\documentclass[aps,prd,twocolumn,superscriptaddress]{revtex4}

\usepackage{graphicx}
\usepackage{dcolumn}
\usepackage{bm}
\renewcommand\({\left(}
\renewcommand\){\right)}
\renewcommand\[{\left[}
\renewcommand\]{\right]}

\newcommand{\ra}{\rightarrow}

\def\lsim{\raise 0.4ex\hbox{$<$}\kern -0.8em\lower 0.62
ex\hbox{$\sim$}}

\def\gsim{\raise 0.4ex\hbox{$>$}\kern -0.7em\lower 0.62
ex\hbox{$\sim$}}

\def\lbar{{\hbox{$\lambda$}\kern -0.7em\raise 0.6ex
\hbox{$-$}}}

\newcommand\eq[1]{eq.~(\ref{#1})}
\newcommand\eqs[2]{eqs.~(\ref{#1}) and (\ref{#2})}
\newcommand\Eq[1]{Eq.~(\ref{#1})}

\newcommand\ee{\end{equation}}
\newcommand\be{\begin{equation}}
\def\bea{\begin{eqnarray}}
\def\eea{\end{eqnarray}}
\newcommand\ees{\end{eqnarray}}
\newcommand\bees{\begin{eqnarray}}

\def\p1{{\bf p}_1}
\def\p2{{\bf p}_2}
\def\k1{{\bf k}_1}
\def\k2{{\bf k}_2}

\newcommand{\hti}{\tilde{h}}

\newcommand{\fmin}{f_{\rm min}}
\newcommand{\fmax}{f_{\rm max}}
\newcommand{\DE}{\D E_{\rm rad}}
\newcommand{\msun}{M_{\odot}}
\newcommand{\rsun}{R_{\odot}}
\newcommand{\ogw}{\o_{\rm gw}}

\def\o{\omega}
\def\O{\Omega}
\def\D{\Delta}
\def\p{\partial}
\def\a{\alpha}
\def\b{\beta}
\def\nn{\nonumber}
\def\th{\theta}
\def\s{\sigma}
\def\g{\gamma}

\def\l{\lambda}

\def\d{\delta}

\def\dslash{\hspace{-1mm}\not{\hbox{\kern-2pt $\partial$}}}
\def\Dslash{\not{\hbox{\kern-4pt $D$}}}
\def\pslash{\not{\hbox{\kern-2.1pt $p$}}}
\def\kslash{\not{\hbox{\kern-2.3pt $k$}}}
\def\qslash{\not{\hbox{\kern-2.3pt $q$}}}

\newcommand{\bt}{\bar{t}}

\begin{document}


\title{On the possible sources of gravitational wave bursts detectable today}


\author{Eugenio Coccia}
\affiliation{University of Rome ``Tor Vergata'' and INFN Gran Sasso}
\author{Florian Dubath}
\affiliation{D\'epartement de Physique Th\'eorique, 
Universit\'e de Gen\`eve, 24 quai Ansermet, CH-1211 Gen\`eve 4}
\author{Michele Maggiore}
\affiliation{D\'epartement de Physique Th\'eorique, 
Universit\'e de Gen\`eve, 24 quai Ansermet, CH-1211 Gen\`eve 4}

\date{\today}

\begin{abstract}
We discuss  the possibility that galactic gravitational 
wave sources might give burst signals at a rate of several events per year, 
detectable by
state-of-the-art  detectors.
We are stimulated  by the results of  the data collected 
by the EXPLORER and NAUTILUS 
bar detectors
in the 2001 run, which suggest an excess  of coincidences between the
two detectors, when the resonant bars are orthogonal 
to the galactic plane. 
Signals due to the coalescence of galactic compact binaries 
fulfill the energy requirements but are problematic
for lack of known candidates with
the necessary merging rate. We examine the limits imposed by
galactic dynamics on the mass loss of the Galaxy due to
GW emission, and we use them to put constraints also on the GW
radiation from  exotic
objects, like binaries made of primordial black holes.
We discuss the  possibility that the events are due
to GW bursts coming repeatedly from a single or a few
compact sources. We examine different possible realizations of this
idea, such as accreting neutron stars,  strange quark stars, 
and the highly magnetized neutron stars
(``magnetars'') introduced to explain  Soft Gamma Repeaters. 
Various possibilities are excluded or appear very unlikely, while others
at present cannot be excluded.

\end{abstract}


\maketitle

\section{Introduction}

In ref.~\cite{ROG01} the ROG collaboration
has presented the analysis of  the data collected 
by the EXPLORER and NAUTILUS resonant bars during nine months in the year
2001. When the number of coincidences between the two resonant bars is
plotted against sidereal time, one finds an excess of
events with respect to the expected background, concentrated around
sidereal hour four. 
At this sidereal hour the two bars are oriented perpendicularly
to the galactic plane, and therefore  their
sensitivity for galactic sources of gravitational waves
(GWs) is maximal.
Furthermore, for these events the energy deposed in one bar is
well correlated with the energy deposed in the other. 
The significance of this observation has been debated
 ~\cite{Finn,ROG02,ROG02bis}, 
and certainly further experimental work will be necessary to put the indications on a
firmer ground. New data from bars and interferometers, from long data taking runs, should 
soon clarify the situation.  

As we will discuss below,
from a theoretical point of view the existence of GW bursts
with the intensity and the rate necessary to explain the 
EXPLORER/NAUTILUS
results, if they will be
confirmed by further data, 
would certainly be a great surprise. It would then be necessary
to reconsider with an open mind a number of unexpected 
possibilities. The purpose of this paper is 
to provide a general
framework for such a study, indicating what are the difficulties, 
which directions of investigation appear to
be more promising and which are ruled out. In particular,  we hope
that our work
will help  to set the stage for the analysis of future data.

The structure of the paper is as follows. In sect.~\ref{sectEnereq}
we present some of  the basic aspects to be explained, that is: (1)  the energy
which must be released in GWs in order to explain the events observed,
and (2) the  rate which is inferred from the EXPLORER/NAUTILUS 
observations. The other crucial experimental information is the
distribution of the events as a function of sidereal time. This is
presented in sect.~\ref{sectSid}, together with a detailed discussion
of what can be learned from sidereal time analysis. We think that
this section has also a more general methodological interest.
In sect.~\ref{sectCoa} we examine one of the most obvious candidates, the
coalescence of a compact binary system. We find that compact objects
of solar masses, at typical galactic distances, would account for the
energy release. However, the rate expected from binaries made of
neutron stars and/or black holes is many order of magnitude smaller
than the observed rate. We then turn our attention to more exotic
binary systems, such as those involving primordial black holes. 
The theoretical uncertainties on exotic events make more difficult to
put direct limits on their rate. However, the coalescence of binary
systems should be a phenomenon which takes place at a more or less
steady rate for a time comparable to the age of the Galaxy. 
We show in sect.~\ref{sectUpper} that
the corresponding loss of mass of the Galaxy into GWs is 
constrained by  galactic dynamics. Again, we think that 
this section is of  more general interest, independently of the
application to the EXPLORER/NAUTILUS data.

These considerations lead us to suggest, in sect.~\ref{bursters}, that
the all events might be generated by a single (or just a few) ``GW
burster'', i.e., by an object which emits repeatedly bursts of GWs, and we
examine different possible implementations of this idea.

In sect.~\ref{sectAcou} we show that the signal cannot be
accounted for by
non-gravitational phenomena like the deposition of recoil energy in
the bars due to the passage of massive particles. We present our
conclusions in sect.~\ref{sectConc}.

\section{Energy requirements}\label{sectEnereq}

Given a GW with Fourier spectra $\tilde{h}_{+,\times}(f)$,
the gravitational energy
radiated  per
unit area and unit frequency
is given by~\cite{Th}\footnote{Occasionally this equation appears in the
  literature with an incorrect factor $1/4$ instead of $1/2$, 
  due to the fact that the
  contribution of the negative frequencies in the plane wave expansion
  has been forgotten. We have checked that the result of
  ref.~\cite{Th} is the correct one.}
\be\label{1dEdAdf} 
\frac{dE}{dAdf}=\frac{\pi c^3}{2G} f^2 
\(  |\hti_{+}(f) |^2 +|\hti_{\times}(f) |^2  \)\, .
\ee
For a wave coming from an arbitrary direction and with arbitrary
polarization, 
a detector  does not measure directly 
$\hti_{+}(f)$ and $\hti_{\times}(f)$ but rather the combination
$\hti (f) =F_+\hti_{+}(f) + F_{\times}\hti_{\times}(f)$,
where $F_{+,\times}$ are the detector pattern function. 
For a bar, $F_+=\sin^2\th\cos 2\psi$, $F_{\times}=\sin^2\th\sin 2\psi$,
where $\psi$ describes the polarizations and $\th$ is the angle of
arrival, measured from the bar axis.
The events we consider here are
recorded when the bar is orthogonal to the galactic plane, and
therefore when $\sin^2\th\simeq 1$. 
Assuming that the source is emitting randomly polarized GWs, we average
over  $\psi$. Then
\be
\langle \hti^2 (f)\rangle \simeq\frac{1}{2}\,
\[
 |\hti_{+}(f) |^2 +|\hti_{\times}(f) |^2
\]\, ,
\ee
where $\langle \ldots \rangle$ denotes the average over the
polarizations and we used $\langle \sin^2 2\psi \rangle=
\langle \cos^2 2\psi \rangle=1/2$ and
$\langle \sin 2\psi \cos 2\psi \rangle=0$.  Therefore \eq{1dEdAdf} becomes
\be\label{1dEdAdf2} 
\frac{dE}{dAdf}=\frac{\pi c^3}{G} f^2 \langle \hti^2 (f)\rangle\, .
\ee
For a burst, we assume that $\langle \hti^2 (f)\rangle$ 
is equal to a constant $\hti_c^2$
between a frequency $\fmin$ and a frequency $\fmax$. Denoting by $r$
the distance to the source and assuming isotropic emission,
 the total radiated energy $\DE$ is therefore given by
\be
\DE = \frac{4\pi^2 r^2 c^3\hti^2_c}{G} \,\, \frac{1}{3}(\fmax^3
-\fmin^3)\, .
\ee
Typically, the term $\fmin^3$ is negligible in comparison
with $\fmax^3$, and
therefore we write simply
\be\label{1DE}
\DE \simeq \frac{4\pi^2 r^2 c^3}{3G} \hti^2_c\, \fmax^3
\, .
\ee
For EXPLORER and NAUTILUS
the value of $\hti_c$ is related to the energy $E_s$ deposed in the
bar by
\be\label{1Eh} 
\hti_c = 2.5\times 10^{-21} \, {\rm Hz}^{-1} 
\( \frac{E_s}{100\,  {\rm mK}} \)^{1/2}\,\, .
\ee
Therefore, under the hypothesis  
that the events recorded  indeed correspond to GWs, 
each burst originates from a process that liberated in GWs the energy
\be\label{1Erad}
\DE \simeq 10^{-2} \msun c^2\, 
\( \frac{E_s}{100\, {\rm mK}} \)
\(\frac{r}{8\, {\rm kpc}}\)^2
\(\frac{\fmax}{1{\rm kHz}}\)^3\hspace{-1mm}.
\ee
A typical value of $E_s$ 
for the considered events is  $E_s\sim 100$~mK for the
2001 data, see ref.~\cite{ROG01}.

In terms of the GW amplitude $h$, again assuming a flat Fourier spectrum
up to a frequency $\fmax$, for a burst of duration
$\D t\simeq 1/\fmax$ we have $h\simeq \hti_c\fmax$, so
\eq{1Eh} gives
\be\label{h2518}
h \simeq 2.5\times 10^{-18} \( \frac{E_s}{100\,  {\rm mK}} \)^{1/2}
\(\frac{\fmax}{1{\rm kHz}}\)\, .
\ee
Assuming that a couple of events are due
to backgrounds, as is expected in a two hours period~\cite{ROG01}, 
we estimate that  the 
8 events recorded correspond to a signal rate of order
$ 200$\, events/yr.
The challenge is therefore to explain what kind of source could give
such a strong GW emission, at such a high rate.
We also remark that there is no
observed  neutrino  counterpart of these events~\cite{Aglietta}.

\section{Sidereal time analysis}\label{sectSid}

The other crucial piece of information  
is the distribution of the number of coincidences when plotted 
against sidereal time, 
as stressed in particular in refs.~\cite{BP1,BP2,BP3}. 
The purpose of this section is to discuss in
detail what we can learn
about the location (and possibly the polarization) of the sources
from the sidereal time analysis.

\begin{figure}
\includegraphics[width=0.45\textwidth, angle=0]{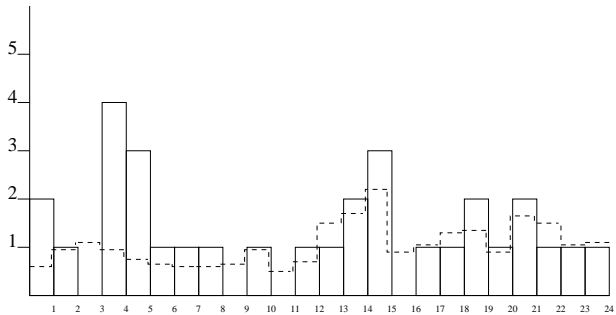}
\caption{\label{figsid1} The number of
coincidences found in the 2001 run (solid
line) and the estimated background (dashed line), 
against sidereal time $\bt=(t_{\rm Expl}+t_{\rm Nau})/2$
(adapted from fig.~5 of ref.~\cite{ROG01}).  }
\end{figure}

The experimental data 
for the 2001 run are shown in
fig.~\ref{figsid1}, adapted from
ref.~\cite{ROG01}. 
The coincidences are binned in one hour bins corresponding to the sidereal
hour of arrival.
The number of
accidental coincidences (dotted line) is estimated shifting the data stream
of one detector 
with respect to the data stream of the other by  a step 
$\d t=2$~s and measuring
 the number of
coincidences, which now are all
accidentals. The analysis is repeated  with a step $2\d t$, then
with $3\d t$, etc., up to $100\d t$.
If $n(j)$ is the number of
coincidences found with a shift $j\d t$,   then the average number of
accidental coincidences is taken to be 
$\bar{n}=(1/100)\sum_{j=1}^{100}n(j)$.
We see from the figure that there is an excess of coincidences with
respect to the expected number of accidentals,
at sidereal hours 3 and 4.
The data in fig.~\ref{figsid1}  refers to events
collected only in periods with more than 12 hr of continuous
operation. If we consider all periods with more than 1 hr of
continuous operation the events at the peak becomes $4+4$ rather that
$4+3$ as in fig.~\ref{figsid1}, and the background also rises
slightly, see fig.~7 of  ref.~\cite{ROG01}.

While the statistical significance of the peak is not large, two
important facts make this result more intriguing: (1) sidereal hour 4
is a very special moment. In fact, a peak at this value of sidereal
time is predicted to appear for sources in the galactic plane.
This is a very general prediction, independent on the precise location 
and nature of the sources.
As we will discuss in  detail below, depending on the distribution
of sources in the galactic plane (e.g., a uniform distribution in the
disk, sources concentrated in the galactic center, etc.)
a second peak, possibly smaller, can appear 
at another value of sidereal time, but the value of $t$ of this second
peak depends instead strongly on
the specific position of the sources. 
(2) For the 8 events at
the peak, the energy deposed in one bar is very well correlated with
the energy deposed in the other, while
for the  events at other sidereal hours the energy deposed in one bar
and in the other are
completely uncorrelated~\cite{ROG01}.

To extract physical informations from the plot against sidereal hour
we proceed as follows.
Let $\th$ denote the angle between the direction of a  source and the
longitudinal axis of the bar. The response 
of a single bar to GWs from
this source is 
\be\label{pola}
\hti (f) =\sin^2\th \( \hti_{+}\cos 2\psi +
\hti_{\times}\sin 2\psi\)\, .
\ee
Therefore the response in amplitude is
proportional to $\sin^2\th$ and,
since the energy is quadratic in the  amplitude, the response of the 
detector in energy is proportional to $\sin^4\th$.
Furthermore, there can be an effect due to the polarization of the
waves, i.e., to the angle
$\psi$. We will  consider first the case of unpolarized GWs and then we
will discuss the possible modification due to the dependence on $\psi$.

\subsection{Randomly polarized GWs}\label{sectunp}

If the GWs come from an ensemble of sources or from a single
source which emits randomly polarized waves, 
the effect of $\psi$ will be averaged out and the
energy deposed in a detector will just be proportional to $\sin^4\th$.
Of course $\th =\th (t)$
because of the rotation of the Earth. Using
a bit of geometry (whose details are left to appendix~A) one can see 
that
\bees
\cos\th (t) &=& [n_x\cos (\a - \O t) +n_y\sin(\a -\O t)] \cos\d +\nn\\
&&+n_z \sin\d\, , \label{theta}
\ees
where
$\a ,\d$ are the right ascension and declination of the
source, $a$ is the azimuth of the bar, $l$ its latitude, 
$n_x =-\cos a \sin l$, $n_y = \sin a$, 
$n_z=\cos a\cos l$,
$\O =2 \pi/24$ is the rotation frequency of the Earth in units $1/$
(sidereal hours), $t$ is the local time
measured in sidereal hours, and all angles, 
including
$\a ,\d$, are expressed in radians. 

The axis of the two detectors are aligned to within a few degrees
of one another, so that the response of the two bars to any source can be
considered identical. The latitudes and azimuths are: 
for EXPLORER, $l=46.45N$ and
$a=39^oE$ while for NAUTILUS $l=41.82N$, $a=44^oE$. The longitudes 
 are 
instead $6.20E$ and $12.67E$, respectively.
Given their difference in longitude, the local sidereal time at the
EXPLORER and at the NAUTILUS locations differ by $\D t \simeq 25.88$
min. We will  denote by $\bt$ the average between 
the two local  sidereal times, $\bt=(t_{\rm Expl}+t_{\rm Nau})/2
\simeq t_{\rm Expl}+0.216$~hr. (The
same variable was used in ref.~\cite{ROG01}.
To compare with other experiments it might be more convenient to use
Greenwich sidereal time $t_{\rm Green}$, related to $\bt$ by
$\bt\simeq t_{\rm Green}+0.629$~hr). 

\begin{figure}
\includegraphics[width=0.4\textwidth, angle=270]{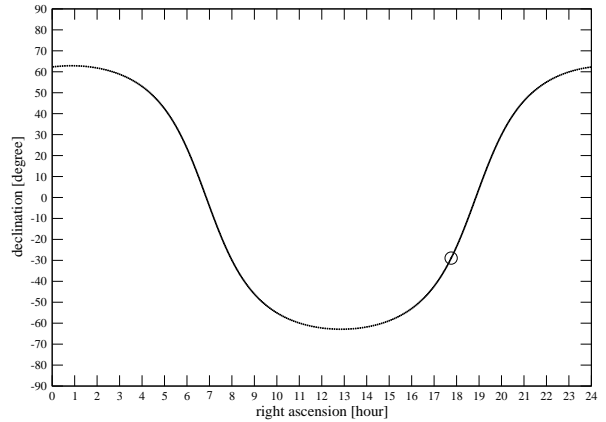}
\caption{\label{galplane} The location of the galactic plane in the
  equatorial coordinate system.
The position of the galactic center is marked by a circle.}
\end{figure} 

In the equatorial coordinate system $(\a ,\d )$ the galactic plane is
represented by the curve in fig.~\ref{galplane}.
For sources in the galactic plane a plot of $\sin^4\th$
against sidereal time $t$, with $0\leq t\leq 24$, always shows two maxima:
one peak 
is close to sidereal hour 4, because at that moment the bar
turns out to be almost
perpendicular to the galactic plane. The precise position of the
source within the galactic plane results only in a minor variation
of the precise
value of $\bt\simeq 4$ where $\sin^2\th=1$. Instead,
the sidereal time at
which there is the
second peak
depends strongly on the  location of the
source.

\begin{figure}
\includegraphics[width=0.3\textwidth]{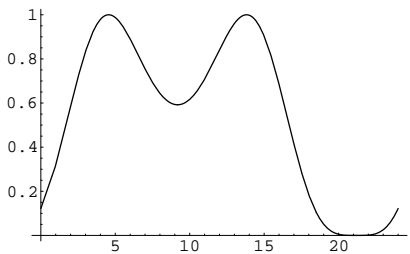}
\caption{\label{figgc} $\sin^4\th$ 
 against sidereal time $\bt$ (in hr) for
  the galactic center.}
\end{figure}

\begin{figure}
\includegraphics[width=0.3\textwidth]{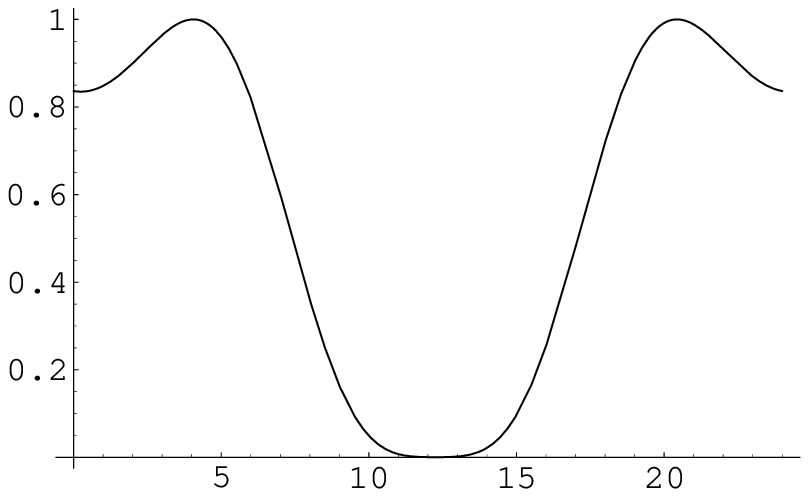}
\caption{\label{figsin4} $\sin^4\th$  against sidereal time $\bt$ (in hr) for a
  source  with coordinates $\a =135^o$ and $\d
  =-40^o$.}
\end{figure}

\begin{figure}
\includegraphics[width=0.3\textwidth]{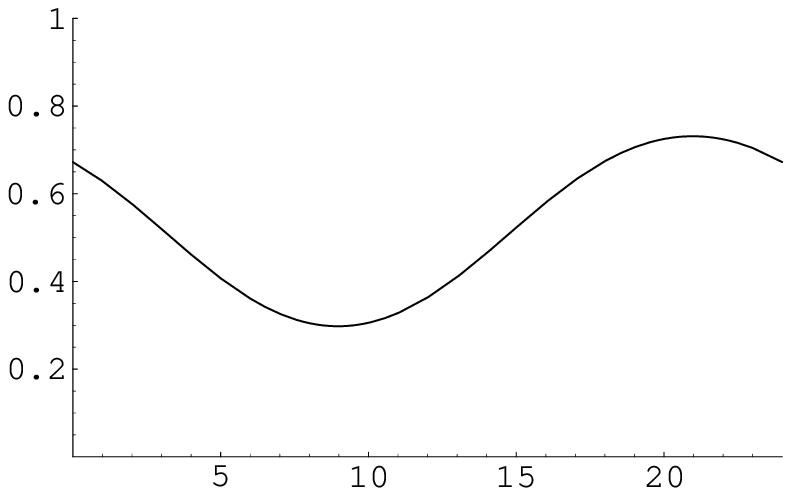}
\caption{\label{fighalo}$\sin^4\th$ against sidereal time $\bt$ (in hr) for a
  source with $\a=266.405^o ,\d = 80^o$.}
\end{figure}

\begin{figure}\hspace*{15mm}
\includegraphics[width=0.4\textwidth, angle=90]{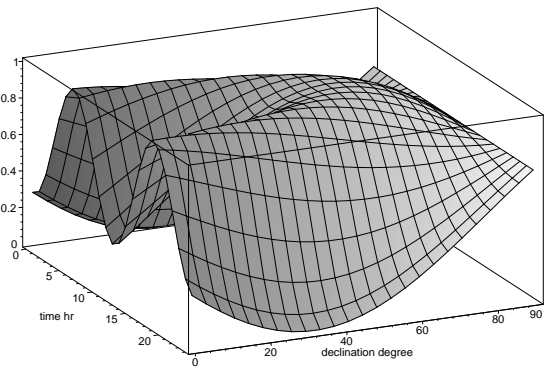}
\caption{\label{sensi}$\sin^2\th$ as a function of
sidereal time $t$ and of the declination $\d$, for $\a = 135^o$.}
\end{figure}

Fig.~\ref{figgc} shows $\sin^4\th$ 
as a function of $\bt$ for a source
located in the galactic center, i.e.,   $\a =266.405^o$
and $ \d =-28.936^o$. In
fig.~\ref{figsin4} we show $\sin^4\th$  for another source in
the galactic plane, taking as an example a source with equatorial
coordinates $(\a =135^o ,\d = -40^o )$.
Both curves have a maximum in the bin  $4<\bt<5$ (more precisely, at $\bt\simeq
4.55$ for the galactic center and $\bt\simeq 4.27$ for the 
source of fig.~\ref{figsin4})
and they have a second peak whose
position depends on the source location. Furthermore, they have  a 
minimum where $\sin^4\th$ becomes zero.
Moving away from
the galactic plane, e.g. increasing the declination
$\d$ at fixed $\a$ (see fig.~\ref{galplane})
at first the curve still has two maxima,
with positions that depend on the values of $(\a ,\d )$, but then
one arrives at a critical value of
$\d$ where the two maxima coalesce. Above this critical value there is
just a single maximum, and
as we increase further
$\d$ the value at the peak  becomes smaller than one. 
At the same time the value at the minimum moves away from zero, and
therefore the response curve becomes 
much flatter.
Fig.~\ref{fighalo} shows the curve for a source with 
$\a=266.405^o ,\d = 80^o$. In
the limit $\d =90^o$ we see from \eq{theta} that 
the curve becomes flat, at a value which depends on $a,l$, 
and which in
our case is 
$\sin^4\th \simeq 0.504$. A plot of $\sin^2\th$ as a function of
the local sidereal time $t$ and of the declination $\d$ is shown in
fig.~\ref{sensi}. The effect of changing $\a$ on this figure is simply
to produce a shift in the origin of sidereal time, as we see from the
fact that in \eq{theta} $\a$ and $t$ appear only in the
combination $\a -\O t$.

\begin{figure}
\includegraphics[width=0.45\textwidth, 
height = 0.32\textwidth, angle=0]{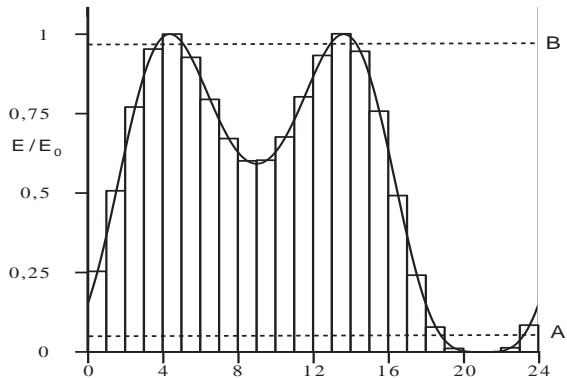}
\caption{\label{figGraph} The energy of the events in the various
  sidereal time bins,
compared to two possible thresholds of the detectors (the dotted lines 
marked as A and B).}
\end{figure}

We now discuss 
how these response curves are reflected 
 in a plot of the number of observed
coincidences, $N_c$, against sidereal
time. Consider  for definiteness the situation in which the sources are
in the direction of the galactic center. 
In fig.~\ref{figgc} we have shown
$\sin^4\th$ as a function of $\bt$ for these sources, and
it is intuitive that the maxima of $N_c$, as a function of $\bt$, 
coincide with the two maxima of $\sin^4\th$, since at that values of
sidereal time the
bars have their best orientation with respect to these sources. However,
it is important to understand that the functional form of
$N_c$, and in particular the width of the maxima,
is not the same as that of the function $\sin^4\th$. 
Rather, it depends crucially on the ratio between 
the energy  of the
events and the energy threshold of the 
detectors~\cite{BP1,BP2,BP3,ROG02bis}.  To understand
this point, suppose that in
a given observation time arrive $n$ events in each one-hour bin and
restrict for the moment to the simplified situation in which
a  signal arriving at a generic sidereal time $t$ 
deposes in the bar an energy $E_0\sin^4\th (t)$, with $E_0$ the same
for all signals.  
Consider first the
limiting case in which the threshold of the detector is very low
with respect to $E_0$, e.g., is represented
by the dotted line $A$ in fig.~\ref{figGraph}. In this case  all
the events are above threshold and therefore are recorded,
except  those falling exactly in the blind direction of
the detectors, which in this example corresponds
to the bins between sidereal hours 19 and
23. Therefore in the bins corresponding to sidereal hours 19 to 23 we would
record zero events, while in all others
we would observe $n$ events
per bin. In this limiting case, therefore, $N_c(\bt )$ shows no peak at all;
rather, it is a flat function, $N_c(\bt )=n$, except for a dip in
correspondence to the
blind direction.

The opposite limiting case is represented by an energy
threshold very close to $E_0$ (the dotted line B
in fig.~\ref{figGraph}). In this case,
only when we are in the two bins where $\sin^4\th$ becomes equal
to one we can
detect the signals, while as soon as we move to the next bin we fall
below the threshold and no signal is detected. In this case 
$N_c=n$ in the bins $4<\bt <5$ and $13<\bt <14$, while $N_c=0$ otherwise.
A plot of
$N_c(\bt )$ will show two very narrow peaks, each one
concentrated just in a single
sidereal hour bin.

In a more realistic situation, the above picture will be complicated by
the fact that the signals will arrive with a distribution in energy,
rather than being monoenergetic. Moreover, the energy of the events is
in any case ``dispersed'' by the fact that the events observed are a
combination of the GW signal and of the noise.
The above example, however, suffices to
illustrate that the width of the peak is crucially affected by the energy
distribution of the events and by the detector thresholds, and can be
very flat (for great values of the
typical signal-to-noise ratio) to very narrow (for small
SNR). 
One should also observe that, 
when the average SNR of the events is small, we can see
$\sin^4\th$ as a measure of the probability that, in a single
detector, a signal is not lost below the energy threshold. In a
two-detector correlation the probability that both detectors see the
signal is therefore rather measured by $\sin^8\th$, which therefore is
a more appropriate envelope for the graph in fig.~\ref{figGraph}, and
this results  in an even  narrower peak.

It is also
interesting to note that, in the case of large SNR, informations 
on the location of the sources cannot be extracted from
$N_c(t)$, which is now basically flat, but it
could be obtained
plotting against sidereal time the average energy of the coincident events,
although probably a  large sample of data would be needed, to
compensate for the spread in the intrinsic energies of the signals.

\begin{figure}
\includegraphics[width=0.45\textwidth, 
height = 0.32\textwidth, angle=0]{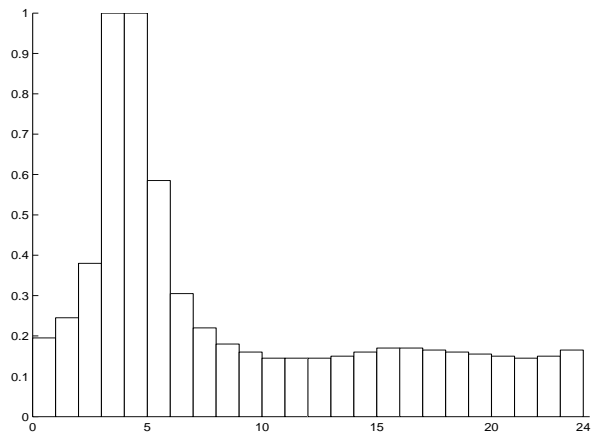}
\caption{\label{fig7bis} A simulation of the number of events per bin,
for a
distribution of sources in the galactic plane (chosen to be
uniform in  galactic longitude) 
when the detector threshold is such that $E/E_0=0.9$.}
\end{figure} 

If we assume a given distribution of sources and a given typical value
of the detector threshold with respect to the signal, we can simulate
the distribution  of the number of events as a function of sidereal time. In 
fig.~\ref{fig7bis} we show an example where, for illustration, we considered 
a population of sources all in the galactic plane, and
distributed uniformly in galactic longitude (this is natural for
sources seen from the Earth, if we can  detect only sources which are
not too far, say $r<0.5$~kpc. 
If one would be sensible also to very far sources, then
clearly toward the galactic center there would be many more sources than in
the opposite direction). We assume that each source deposit in the bar
an energy
$E_0\sin^4\th (t)$, with $E_0$ the same for all bursts and for all sources. 
We fix a detector threshold (in fig.~\ref{fig7bis} we have chosen a
rather high threshold, $E/E_0=0.9$). For each source we produce a
plot as fig.~\ref{figGraph}, and we see which bins are above
threshold. Then as the contribution of this source we take
 $N_c(t)=1$ for these bins and $N_c(t)=0$ otherwise.
Summing the contributions of all sources and normalizing the
distribution so that the value at the peak is one, we get
fig.~\ref{fig7bis}. Of course this is a rather simplified model, but
it illustrates that a distribution of galactic
sources may produce a single peak
close to sidereal hour four (whose width is controlled by the threshold
that we set).

Comparing with the experimental data in fig.~\ref{figsid1} we see that
the preliminary indication, within the  statistics
 available,  is that there could indeed be a peak  near
sidereal hour 4. No second peak of some statistical significance can be seen.
From the above discussion, this 
points toward the existence
of several sources distributed across the galactic disk. 
Independently of the details of the models considered in
fig.~\ref{fig7bis}, the result follows more generally from the fact that
for a distribution of sources in the galactic plane
each single source would contribute
either to the bins around  sidereal hour 4
common to all sources in the galactic plane, or to
another bin which depends on the specific source location. Only in the
former case the contributions from different sources
add up and, within the present
statistics, give rise to something which emerges over the background.

However, here we should not rush toward conclusions. Even assuming
that the events really correspond to GWs, it is clear that
the full form of
the curve can  be understood only when sufficient statistics will be
available. We will therefore keep open, for the moment, the
possibility that further structures in the sidereal time plot 
might appear with further data, and we will examine different
possibilities. 

We also mention that the sidereal time analysis can be very useful
even in the
study of extragalactic sources~\cite{BP1,BP2}. However, \eq{1Erad}
clearly excludes an interpretation in terms of GWs of extragalactic
origin for the EXPLORER/NAUTILUS data, and therefore 
we will focus on galactic sources.

\subsection{Polarized GWs}\label{sectPolGW}

In sect.~\ref{bursters} we will examine the possibility that all
events come from just one  source which emits repeatedly GW bursts. 
One can expect from such a source a coherent motion of matter 
with some given quadrupolar pattern
related to the source geometry and to
its rotation axes, and consequently the emission of polarized GWs. 
As we will see in this section, the polarization
can affect the sidereal time analysis in an interesting way.

The response of the detector (in amplitude) is given in \eq{pola}, and
the polarization angle $\psi$ 
depends on sidereal time, just
as $\th$. The calculation of the dependence of this
response function on local
sidereal time $t$, for a given location of the source and a given
choice of the polarization axes, is in principle straightforward but 
somewhat lengthy, and we give the details  in Appendix~A.

In fig.~\ref{figpol1} we show the result for 
a source located in the
direction of the galactic center (a similar result is reported in
ref.~\cite{BGMP}). In this figure the envelope is 
$\sin^4\th$, i.e. the response
function (in energy) for an unpolarized source, and is therefore the
same quantity already shown in fig.~\ref{figgc}. The thick line is the
function $\sin^4\th\cos^2 2\psi$, i.e.  the response
function (in energy) for a source which has only the $+$ polarization,
while  the dotted curve is $\sin^4\th\sin^2 2\psi$, i.e. the
response function  in energy for the $\times$ polarization. (The
angle $\psi$ is measured with respect to an axis orthogonal to the
propagation direction and lying along the galactic plane, see 
Appendix~A).

\begin{figure}\hspace*{25mm}
\includegraphics[width=0.4\textwidth, angle=90]{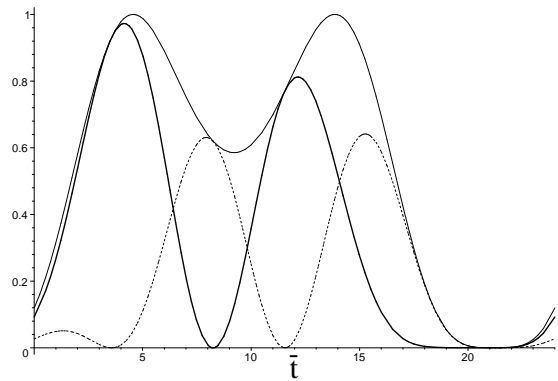}
\caption{\label{figpol1} The response function in energy for unpolarized GWs
  (the envelope), for the $+$ polarization (thick line), and for the
  $\times$ polarization (dotted line), for a source at the Galactic center.  }
\end{figure}

From this figure we see  a number of interesting effects. First of all,  the
maxima are shifted with respect to the unpolarized case.
This is due to the fact that the maxima of $\sin^22\psi$ and of
$\cos^22\psi$ do
not coincide in general with the maxima of $\sin^4\th$, and therefore
the position of the peaks in the response function is determined by a
combination of the two effects. For the same reason, 
the values of the response function at 
the  peaks will be smaller than one, and
need not be equal among the two peaks. We also see that the peak close
to sidereal hour 4 in the $+$ polarization is
 narrower compared to the unpolarized case,
because for the $+$ polarization the maximum of  $\cos^2 2\psi$ is
close to the maximum of $\sin^4\th$ (as we see from the fact 
that the position of the peak close to sidereal hour 4
is only slightly shifted) and therefore 
the additional factor $\cos^2 2\psi$
 in the 
energy response gives a further suppression outside the peak. New
blind directions appear, when $\cos^22\psi=0$ or $\sin^22\psi=0$,
which contribute to make the peaks narrower.

\begin{figure}\hspace*{20mm}
\includegraphics[width=0.4\textwidth, angle=90]{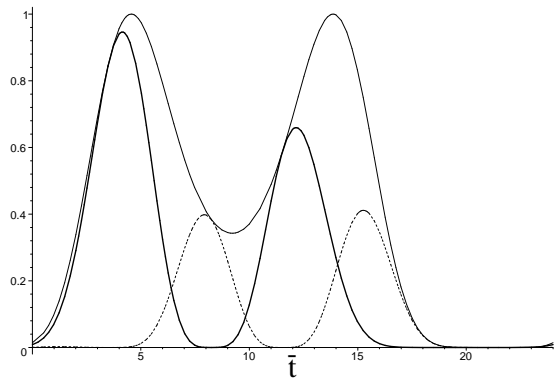}
\caption{\label{figpol2} The square of the
quantities shown in fig.~\ref{figpol1}
 (which is the appropriate response function for a two-detector correlation).}
\end{figure} 

Recalling that the response function of a two-detector correlation,
for low typical SNR, is
the square of the single-detector response function, 
we see that  finally we can obtain  very narrow peaks,
as  shown in fig.~\ref{figpol2}. 
This means that in the case of a source emitting waves 
with a high degree of polarization
along the galactic plane
it is not necessary to have a very low value of the
typical SNR in order to find a narrow peak in the number of observed
coincidences $N_c(\bt )$. A threshold of order $0.5$ in the response
function of fig.~\ref{figpol2} suffices to have $N_c(\bt )$ concentrated
in just two hour bins around $\bt =4$, for a source radiating the $+$
polarization. 
Therefore the number of observed events is
not necessarily a large underestimate of the total number of events, as
it would instead be the case if most of the signals were lost below
the threshold.

\begin{figure}\hspace*{20mm}
\includegraphics[width=0.35\textwidth, angle=90]{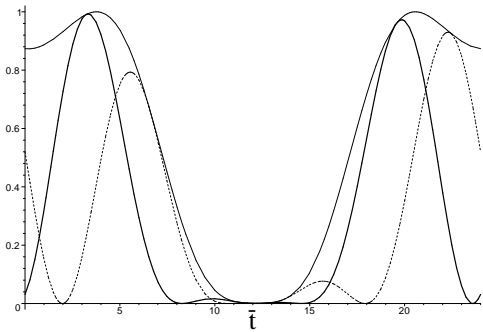}
\caption{\label{figVelPol} The same as fig.~\ref{figpol1},
 for a source with coordinates $\a =135^o, \d =-40^o$, and with a shift 
 $\psi \ra \psi -\psi_0$ with $\psi_0=0.25$ rad.  }
\end{figure}

In this example we have considered the (rather special) case in which
the $+$ polarization is perpendicular to the galactic plane, i.e., the
two axes used to define $h_+$ and $h_{\times}$ are  one parallel and
the other perpendicular to the galactic plane (see Appendix~A). More
in general, we can rotate these axes in the transverse plane by an
angle $\psi_0$. A source which, with respect to these new axes,
emits purely the $h_+$ (or purely the
$h_{\times}$) polarization will have a response function 
obtained shifting $\psi\ra\psi-\psi_0$,
and therefore the position of the
peaks depends on $\psi_0$. As an example, we show
in fig.~\ref{figVelPol} the response to the $+$ and $\times$
polarization, 
for a source located at $\a =135^o, \d =-40^o$ (i.e., the same source as
in fig.~\ref{figsin4}) and with $\psi_0=0.25$ rad.

\section{Coalescence of compact binaries}\label{sectCoa}

As we will recall
below, for  binaries with neutron stars (NS)
and/or black holes (BH) the expected galactic merging
rates are far too small, compared
to  the value $O(200)$ events per year discussed in sect.~\ref{sectEnereq}.
However,
stimulated by the experimental data, one might wish to consider more
unusual possibilities, like BHs of primordial origin or other exotic
objects. For this reason, and also in order to understand better the
difficulties of finding a good candidate source for the
EXPLORER/NAUTILUS data, we start our discussion with 
the coalescence of compact binaries.

The merging of NS-NS, NS-BH or BH-BH binaries
is a  process that  generates GWs
at the kHz, where the bars are sensitive.
Furthermore, they typically liberate an energy of order
$10^{-2}\msun c^2$  in GWs. This can be estimated
first of all analytically, using the wave form of the inspiral phase,
\bees\label{1plus}
h_+(t) &=& -\frac{2^{4/3}}{r}(GM_c)^{5/3}\ogw^{2/3}(\tau )\, 
\( \frac{1+\cos^2\iota}{2}\) \cos [\Phi (\tau ) ]\, ,\nonumber\\
h_{\times}(t) &=& -\frac{2^{4/3}}{r}(GM_c)^{5/3}\ogw^{2/3}(\tau )\, 
 \cos\iota\, \sin [\Phi (\tau )]\, ,
\label{1times}
\ees
where $M_c$ is the chirp mass, 
$M_c= (M_1M_2)^{3/5}/(M_1+M_2)^{1/5}$, $M_1,M_2$ are the masses of the
two objects, $\iota$ is the inclination of
the orbit with respect to the line of sight,
$\tau$ is the time to coalescence,
$\Phi$ is  the accumulated phase and $\ogw (\tau )$ is the
chirping frequency of the GW. Explicitly,
\be
\Phi (t) = \frac{8}{5} \(\frac{5^{3/8}}{4}\) (GM_c)^{-5/8}\tau^{5/8}
+\Phi_0\, ,
\ee
\be\label{1binomega}
\ogw (\tau ) =  \(\frac{5^{3/8}}{4}\) (GM_c)^{-5/8}\tau^{-3/8}\, .
\ee
Taking the Fourier transform in the saddle point approximation (see e.g.
ref.~\cite{Cut}) one gets, apart from irrelevant phases,
\bees
\hti_{+}(f)&=&
 \frac{1}{ \pi^{2/3}}\,\(\frac{5}{24}\)^{1/2}
\frac{1}{r} \frac{(GM_c)^{5/6}}{f^{7/6}}\, 
\( \frac{1+\cos^2\iota}{2}\)\, ,\label{1htichirp}\nonumber\\
\hti_{\times}(f)&= & 
 \frac{1}{\pi^{2/3}}\,\(\frac{5}{24}\)^{1/2}
\frac{1}{r} \frac{(GM_c)^{5/6}}{f^{7/6}}\, \label{1htichirp2}
\cos\iota\, .
\ees
Using \eq{1dEdAdf} 
and performing the angular integration
one finds the energy spectrum,
\be
\frac{dE}{df} = \frac{\pi^{2/3}}{3G}\, (GM_c)^{5/3}\, f^{-1/3}\, .
\ee
Integrating up to a maximum frequency $\fmax$ for which we are still
in the inspiral phase we can estimate the
total energy radiated during the inspiral phase,
\be\label{1Deltachirp}
\Delta E_{\rm rad} \simeq  \frac{\pi^{2/3}}{2G}(GM_c)^{5/3}\, 
f_{\rm max}^{2/3}\, ,
\ee
or, inserting the numerical values,
\be\label{1Eradchirp}
\Delta E_{\rm rad}\simeq 4.2\times 10^{-2} \msun c^2 \, 
\(\frac{M_c}{1.22\msun}\)^{5/3} \,
\(\frac{\fmax}{1\,{\rm kHz}}\)^{2/3}\, ,
\ee
where the normalization of the chirp mass corresponds to
$M_1=M_2=1.4\msun$.

This simple analytic 
treatment gives a value in very good agreement with that found
numerically, from  sophisticated hydrodynamical simulations of NS-NS
coalescence including both the inspiral and plunge phases~\cite{Ruf},
\be\label{merge}
\DE\simeq (1-3)\times 10^{-2}\,\msun c^2\, .
\ee

\begin{figure}
\includegraphics[width=0.45\textwidth]{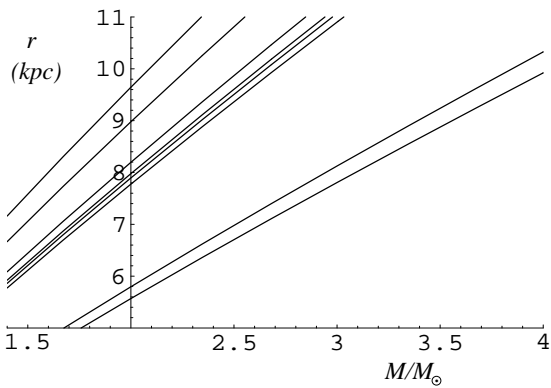}
\caption{\label{fig1}The distribution of distances $r$ against mass
  $M$ for the eight 2001 events at the peak at sidereal hours 3 and 4.}
\end{figure}
\begin{figure}
\includegraphics[width=0.45\textwidth]{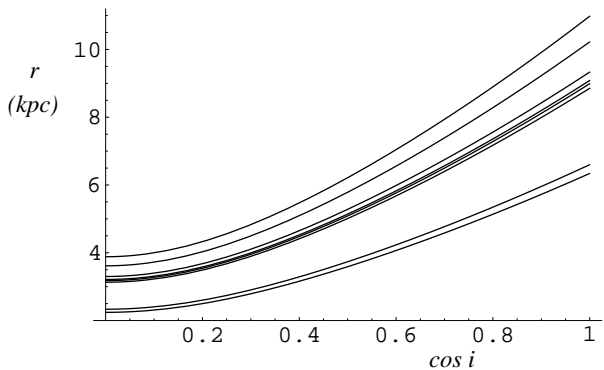}
\caption{\label{fig2}The distribution of distances $r$ against $\cos\iota$
  for the eight 2001 events at the peak at sidereal hours 3 and 4, setting
  $M=1.35\msun$.}
\end{figure}

Comparison with \eq{1Erad} indicates that, from the energetic point of
view, the coalescence of compact binaries is an interesting candidate.
Actually, \eq{1Erad} has been obtained assuming, for lack of better
informations, a square waveform in frequency
space. Since in the
case of an inspiral we have the exact waveform, we can be more
precise. Averaging over the angle $\psi$ we have in fact
\be
\hti_c^2=\frac{1}{2}\sin^4\th \( |\hti_{+}(f) |^2 +|\hti_{\times}(f)
|^2\) \, ,
\ee
while from \eq{1htichirp2} we have
\be\label{1Both}
|\hti_{+}(f) |^2 +|\hti_{\times}(f) |^2 =
\frac{5}{24\,\pi^{4/3} r^2}
\frac{(GM_c)^{5/3}}{f^{7/3}}\,  g(\iota )\, ,
\ee
where
\be
g(\iota ) = \( \frac{1+\cos^2\iota}{2}\)^2 +\cos^2\iota\, .
\ee
We can then substitute this  expression for $\hti_c$ into \eq{1Eh}. 
We set $f\simeq 920$~Hz (corresponding to the
resonant frequency of the bars). For events coming from the galactic plane,
detected around sidereal hour four, we can also set $\sin\th\simeq 1$.
Then
 we find
\be\label{3eug}
\frac{E_s}{100\,{\rm mK}}\simeq 1.1 \, \(\frac{M_c}{1.22\msun}\)^{5/3}\,
\(\frac{8\,{\rm kpc}}{r}\)^2\, \(\frac{g(\iota )}{0.8}\)\, .
\ee
We have chosen as a reference value for $g(\iota )$ its average value
over the solid angle.  
Each event, i.e. each value of $E_s$, determines  a curve in
the plane $(\tilde{M}_c, r)$, where
\be
\tilde{M}_c = M_c \,\( \frac{g(\iota )}{0.8}\)^{3/5}
\ee
and this allows
us to check whether each event can be interpreted
with reasonable  values for $M_c$ and $r$.
The function $g(\iota )$ varies between
between $1/4$ for $\cos\iota =0$, i.e. when the observer is in the
plane of the orbit, 
and $g(\iota )=2$ for $\cos\iota =1$, i.e. when the line of sight of
the observer is perpendicular to
plane of the orbit. Correspondingly, $\tilde{M}_c$ ranges between 
$0.50M_c$ and $1.73M_c$.

The results are shown in fig.~\ref{fig1}, for the eight events 
at sidereal hours 3 and 4 in the 2001 data.  We
  have written $M_c=M/2^{1/5}$, which is the chirp mass of
 a  system with two
  equal masses   $M_1=M_2=M$, and we have fixed $g(\iota )=0.8$.
In fig.~\ref{fig2} we fix instead
  $M=1.35\msun$, a typical value for a NS, and we plot $r$ against
  $\cos\iota$ for the same eight events.

We see that very natural  values of $M$ and $r$ are obtained. Binary
systems composed of compact solar-mass objects, at typical galactic distances,
would be compatible with the energetic observed in the events.

The problem with these sources, however, is the rate.
Before the recent discovery of a new  NS-NS
binary,
the rate of  NS-NS coalescences in the Galaxy
was estimated
to be in the range $10^{-6}$ to $5\times 10^{-4}$
mergings per year~\cite{KNST}. 
This estimate depends strongly on the shortest-lived systems known, and
the recent discovery of a new  NS-NS
binary, with the shortest known merging time (85 Myr),
brings this estimate up by one order of magnitude,  possibly up to a
factor 30~\cite{Bur}; still, we
are  very far from the $O(200)$ events per year needed to
explain the NAUTILUS/EXPLORER observations.
The estimates of the
coalescence rate of binary systems involving black holes (BH-BH
and BH-NS) can be based only on stellar evolution models, rather than
on observations, since no such systems have yet been observed. The
theoretical calculations suffers from large uncertainties, but 
the rate  for BH-BH and BH-NS coalescence is estimated to be
of the same order of magnitude (or 
smaller) than the NS-NS coalescence rate, 
see e.g. ref.~\cite{BKB} and references therein.

Under the stimulus of the NAUTILUS/EXPLORER data, one might 
consider more exotic possibilities, as for example black hole of primordial
rather than astrophysical origin. In this case the estimates based on
stellar evolution do not apply, and it is
interesting to observe that primordial BHs produced in the early
universe at the QCD phase transition would have today a mass of approximately
$0.5\msun$~\cite{Jed}.  The density of MACHO's, and therefore also
of primordial BHs of mass $\sim
0.5\msun$, can be constrained from the microlensing of stars from the
Large Magellanic Cloud~\cite{Alc,Benn}. These data suggest that the
fraction of our Galaxy's stellar mass  which is in the form of black
holes can be significantly larger than 1\%.
With some assumptions, the fraction of primordial black holes in
binary systems, and their distributions in eccentricity and semi-major
axis has been estimated in \cite{NakTh,Nak}, and it has been found that
the coalescence rate of primordial black hole binaries  could be of order
$5\times 10^{-2}$ events per year per galaxy. This is
still far too small to explain the data. 
Furthermore, a stellar  population of
primordial origin is expected to reside in the galactic halo rather than in
the galactic disk, so
 there should be no reason to get an enhanced
signal when the bars are oriented favorably with respect to the
galactic plane.
However, when one enters into the realm of exotic possibilities
there are large uncertainties in the theoretical estimates. 
For instance, for primordial black holes, there are large uncertainties
on the detailed formation mechanism, and therefore one should keep an
open mind on this option.

Other interesting informations might come observing that the
coalescence of any population of compact binaries should be a phenomenon
which  takes place at a more or less
steady rate for a timescale comparable to the age of
the Galaxy. It is therefore useful to examine whether, independently
of the nature of the sources, 
the corresponding mass loss of the Galaxy to
GWs is compatible with known facts of galactic dynamics. We discuss
this issue in the next section.

\section{Upper limits on the GW emission from the Galaxy}\label{sectUpper}

From the energy radiated in each burst, \eq{1Erad}, and a rate
$O(200)$ events/yr, we see that the mass lost by the Galaxy into GWs
is of order
\be\label{1yr}
-\dot{M} \sim 2 \frac{\msun}{\rm yr}\, 
\(\frac{r}{8\, {\rm kpc}}\)^2
\(\frac{\fmax}{1{\rm kHz}}\)^3\, ,
\ee
where now $r$ is an average distance of these sources.
If the events are indeed due to the merging of compact objects we know the
energy which has been released by each burst, independently of 
the values of $r$
and of $\fmax$, 
see 
\eq{merge}, and therefore we can write
\be\label{310}
-\dot{M} \sim (2- 6)  \frac{\msun}{\rm yr}\, .
\ee
This is a very large mass loss. For comparison, the mass loss of
 the Galaxy due to electromagnetic radiation is 
$9\times 10^{-3} \msun/{\rm yr}$. In this section we will
investigate whether such a high value is compatible with known facts of
galactic dynamics.

If the mass loss is really due to GWs from a number of different
sources distributed across the 
galactic disk, it is difficult
to imagine that its rate changed dramatically with time
over the age of the Galaxy. Assuming a
steady mass loss at the level of \eq{310} for a time comparable
to the age of the
oldest clusters in the Galaxy, $T\sim 1.2\times 10^{10}$~yr, gives a
total mass loss over the history of the Galaxy of at least
$\D M\sim  2\times 10^{10}\msun$. For comparison,
the total mass of the disk is estimated to be $6\times 10^{10}\msun$ and the total
mass of the disk plus bulge and
spheroid is  $9\times 10^{10}\msun$. The total
halo mass is $M_{\rm halo}\sim 2\times 10^{12}\msun$ \cite{WE},
but for us this is a less
relevant reference value, since this is the mass
in a spherical halo with radius of order
170 kpc, and therefore events coming from the halo would  not show
any special correlation with the galactic plane.

Of course, the present value of the mass of the
Galaxy gives only a first scale for comparison, since on the one hand
we can imagine that
the mass of the Galaxy could have been  larger
at the formation and, on the other hand, tighter constraints on the
possible mass loss may come from
galactic dynamics. Some of these issues have been considered many
years ago~\cite{SFR,PA,OB}, and it is
interesting to go back to these considerations using the present
knowledge of galactic dynamics, and to compare 
with the energy loss given in eq.~(\ref{310}). 
The most stringent limits are
discussed in the following subsections.

\subsection{Effect of the mass loss on the radial velocity of
  stars}\label{ss1} 

Let us first recall the  basic fact of galactic dynamics that, because
of the differential rotation of the Galaxy, the  radial velocities $v_r$
of stars as seen from the sun, at first order in the distance $R$
between the sun and the star  (after correcting for the solar motion
and averaging over the peculiar velocities of the individual stars)
is given by
\be\label{ur1}
v_r = A R \sin 2l\, ,
\ee
where $A=(-1/2) (Rd\o /dR)_{\odot}$ is Oort's $A$ constant, $l$ is the
galactic longitude, and for simplicity we have written only the expression
valid for  galactic latitude $b=0$ (see e.g.~\cite{BT}).

If the Galaxy is loosing mass, stars become less and less bound and
acquire radial velocities with respect to the Galaxy rest frame.
The physics of the effect can be understood easily 
limiting ourselves to  a star in a circular orbit of radius $r$ in the
galactic plane, 
with a Keplerian longitudinal velocity $v_l$ given by $v_l^2=GM/r$;
$M$ is  an effective mass related to the mass at the
interior of $r$. In particular the effective mass inside the solar circle
is obtained from $v_l\simeq 220$~km/s and $r\simeq 8$~kpc and is
$M\simeq 8\times 10^{10}\msun$.
The conservation of the
angular momentum $J=mrv_l$ gives $\dot{r}v_l+r\dot{v}_l=0$, or
$v_r/r =-\dot{v}_l/v_l$, where $v_r=\dot{r}$ is the radial velocity
in the frame of the Galaxy. Using the
Keplerian value for $v_l$, this gives
\be\label{vr}
v_r = -\frac{\dot{M}}{M}\, r\, .
\ee
In the frame of the sun, this gives an additional contribution to the
radial velocity (\ref{ur1}),
\be\label{ur2}
v_r = A R \sin 2l +K R\, ,
\ee
where 
\be\label{K}
K=-\frac{\dot{M}}{M}\, .
\ee
and $R$ is the sun-star distance.
In the literature on galactic dynamics the additional term 
in \eq{ur2} is known as a K-term. (More generally, one can 
define a K-term as any additional term to be added to 
the right-hand side of \eq{ur1}, independent of the galactic
longitude, and with an arbitrary dependence on $R$, e.g. a constant.
We see that
in the case of mass loss one gets a K-term
growing linearly with $R$). 
The difficulty of extracting from the data the effect
due to the mass loss is that the K term receives
contributions from many other effects of galactic dynamics; 
for example, the value of $K$
extracted from young stars at distances $R<0.6$~kpc is strongly
influenced by the kinematic peculiarities of the Gould Belt, and
therefore the value of $K$ depends on the distance of the sample, and
it turns out to depend also on the age of the stars chosen. To have an
idea of the systematics involved,
we observe that ref.~\cite{TFF}, using the 
Hipparcos data, finds that for $0.1 {\rm kpc} \leq R\leq 0.6$~kpc,
$K$ (expressed in km/(s kpc) )  
ranges from $7.1\pm 1.4$ if one uses stars younger than 30~Myr to 
$-5.4\pm 2.3$ for stars with age between 60-90~Myr, with an average over all
stellar ages  $K=0.5\pm 0.9$. Samples including only stars at larger
distances, and therefore 
insensitive to the Gould belt, give instead negative
values of $K$; in particular, including only stars 
with  $0.6 {~\rm kpc} \leq R\leq 2$~kpc, 
and averaging over all stellar ages, one finds $K=-2.9\pm 0.6$~\cite{TFF}.
A negative value of $K$ might in principle be due to the influence of
the spiral arm structure, but even taking into account spiral arm
kinematics, the sample at $R>0.6$~kpc gives a negative value,
$K=-(1-3)\, {\rm km}\,\, {\rm s}^{-1}\, {\rm kpc}^{-1}$~\cite{FFT}, and
the physical origin of the negative sign, representing a contraction
rather than an expansion, is not really well understood.

To get a bound on the energy loss to GWs, one should 
in principle  be able first of all
to understand the other mechanisms which are at work
and which give a negative contribution to $K$, evaluate them
precisely, subtract them,
and see if one is left with a positive value of $K$ growing linearly
with $R$. This is beyond the accuracy of our knowledge of the Galaxy.
However, in any case we
certainly do not expect a fine tuned cancellation between some
negative contributions to $K$ and the positive contribution due to a
possible GW emission. Let us assume for definiteness that a positive
contribution to $K$ from GW mass loss is smaller  than
20\% of the absolute value of $K$.
We use the value measured at $R>0.6$~kpc, to get rid of effects due to
the Gould Belt, i.e.,
$K=-(1-3)\, {\rm km}\,\, {\rm s}^{-1}\, {\rm kpc}^{-1}$, 
and therefore we   assume 
that the positive contribution from GWs is smaller than
$O(0.4)\, {\rm km}\, {\rm s}^{-1}\, {\rm kpc}^{-1}$. From
\eq{K}, using $M=8\times 10^{10}\msun$, this translates into
\be\label{bound1}
-\dot{M}< O(30) \msun /{\rm yr}\, .
\ee
Clearly, the precise value of the bound can be modified somehow,
changing the level of fine tuning that one is willing to tolerate, but
\eq{bound1} gives a first order-of-magnitude estimate.

\subsection{Mass loss and outward motion of the LSR}\label{ss2}

Rather than looking at the $K$ term, i.e. at
the expansion/contraction of the stars within a few kpc from the sun,
one can  investigate whether the local standard of rest (LSR) has an
overall outward radial velocity, as suggested by \eq{vr}. Long ago
Kerr~\cite{Kerr} indeed suggested an outward radial 
motion of the LSR with a velocity $u_{\rm LSR} =7$~km/s,
as an explanation for the lack of axisymmetry 
of the galactic rotation curve obtained with 21 cm surveys,
and this effect was attributed to
mass loss due to GWs in ref.~\cite{SFR}.  In more recent years 
the experimental determination of $u_{\rm LSR}$ has not become much
more clear. Blitz and Spergel~\cite{BS} from 21 cm line emission
find $u_{\rm LSR}=+14$~km/s (where
the positive sign means radially outward); results consistent with
this value have been found from Cepheid kinematics~\cite{MCS}
while
in ref.~\cite{KT}, from a variety of measurements (OH/IR stars,
globular clusters, high-velocity stars, planetary nebulae), 
it is proposed
$u_{\rm LSR}=-1\pm 9$~km/s. From young open clusters~\cite{Hron} one finds
a maximum value $u_{\rm LSR}=3\pm 2$~km/s.
Recent work on OH/IR stars~\cite{DGS} gives
$u_{\rm LSR}=2.7 \pm 6.8$~km/s. 

However, even if an outward
velocity of the LSR of a few km/s indeed exists, it cannot be due
to mass loss. In fact, the  observation of the
21 cm absorption line toward the galactic center~\cite{RS} shows
that the gas along the line-of-sight  has a mean 
radial velocity with respect to the LSR of  
$-0.23\pm 0.06$ km/s. The absorbing material is probably at 1-2 kpc from the
galactic center. A radial expansion due to mass loss predicts a radial velocity
$v_r\sim r$, \eq{vr}, and therefore, if at the sun location $r\simeq
8$~kpc this effect has to
be responsible for a velocity of the LSR,
we should see a comparable difference in velocity
between us and this gas, $\D v_r =(-\dot{M}/M)\D r$.
The model of Blitz and Spergel  based on 
the existence of a rotating triaxial spheroid~\cite{BS} manages to
escape this limit because it predicts a dependence of $v_r$ on $r$
rather flat between 3 and 8 kpc from the galactic center.
For a mass-loss model instead $v_r\sim r$ and
we do not have this escape route.

As in the previous section, 
there can be in general both positive
and negative contributions from different physical mechanisms
to the value $\D v_r =-0.23\pm 0.06$ km/s, and to
extract a bound on mass loss to GW (or, for that matter, to any mass
loss) we require that no fine tuning between different contributions
takes place. Again we set conventionally  at 20\% the maximum fine
tuning that we allow, which means that
we say that a positive contribution from GWs
to $\D v_r$, if it exists at all, 
must be smaller than $ O(0.04)$~km/s. Setting the distance
between us and the gas to $\D r= 6$~kpc, this gives a bound on the
mass loss to GWs, 
$(-\dot{M})/{M} < O(0.04)\, {\rm km}\,  {\rm s}^{-1}/{6\, {\rm kpc}}$;
using again $M=8\times 10^{10}$, this means
\be\label{005}
-\dot{M}< O(0.5)\,  \msun/{\rm yr}\, .
\ee
Again, we should repeat the caveat that this value depends 
somehow on the
level of fine tuning that we allow but it is clear that to stretch
it to, say, $10\msun/{\rm yr}$, we must make rather unnatural
fine-tuning assumptions.

\subsection{Upper limits from globular clusters and wide binaries}
\label{ss3}

Upper limits of the same order of magnitudes have been found
by Poveda and Allen~\cite{PA} using globular clusters as probes. The
idea is that, if the mass of the Galaxy was much bigger in
the past, the orbits of globular clusters would have been much closer
to the galactic nucleus, and this close interaction with a very
massive central nucleus would have produced the tidal disruption of
the  cluster. Using in particular the globular cluster 
Omega Centauri (NGC5139) as a probe, and assuming a mass loss
localized in the galactic center, ref.~\cite{PA} finds that the 
fact that this cluster still exists today
implies that the Galaxy cannot have lost
more than $O(1)\times 10^{10}\msun$ over its lifetime, implying a bound
on a steady mass loss
\be\label{OmCen}
-\dot{M}< O(1)\,  \msun/{\rm yr}\, .
\ee
Actually, Omega Centauri is a rather unusual globular cluster because
its metalicity distribution, mass and shape lend support to the idea
that it is the remain of a dwarf galaxy which has been tidally
stripped by our Galaxy~\cite{Maj,Tsu}. In this case the considerations of
ref.~\cite{PA} would not apply, since they assume  an adiabatic
evolution of the orbit. For this reason,  the analysis
of ref.~\cite{PA} has been repeated in ref.~\cite{DR}
for two more globular clusters,  M92 and M5. For M92
one finds basically the same limit as for  Omega Centauri, while M5
gives a limit larger by a factor of two.

It should be observed, however, that \eq{OmCen} is obtained assuming a
mass loss concentrated in the galactic center. Since the motion of the
cluster is sensitive only to the mass in the interior of the orbit, if
the mass loss is distributed uniformly over the galactic disk, the
limit is relaxed by an order of magnitude, and one finds~\cite{DR}
\be
-\dot{M}< O(10)\,  \msun/{\rm yr}\, ,
\ee
from $\o$-Cen and M92, and 
$-\dot{M}< O(20)\,  \msun/{\rm yr}$ from M5.

A limit comes also from the existence of old wide binaries, since for
a very massive galactic nucleus  the galactic orbits would have
been much smaller than at present, therefore the density of stars
would have been much larger and the dissolution time of binaries due
to stellar encounters correspondingly shorter. From a list of 11 
well observed old wide binaries Poveda and Allen~\cite{PA} find a
limit on steady mass loss
\be
-\dot{M}< O(10)\,  \msun/{\rm yr}\, .
\ee

\subsection{Comparison with the Explorer/Nautilus data}

Comparing the limits discussed in these sections with the mass loss 
given in eqs.~(\ref{1yr}) or (\ref{310})
we can make the following considerations. 
Taking into account the uncertainties in the theoretical bounds, 
we cannot exclude  that GW emission
can generate a steady
mass loss at the level of \eq{1yr} or \eq{310}. However, such a
large mass loss can be reconciled with the bounds discussed only  invoking
a certain amount of fine tuning.
It should also be mentioned that the galactic
disk accretes external matter, so this could partially compensate the
mass loss. However,  the accretion is mostly from the edge of the
disk and in this case it should not influence much our considerations.

We can therefore summarize as follows the points which must be
addressed in order to advance an
interpretation of the data in terms of a population of sources residing in the
galactic disk. (i) The estimated merging rates of NS-NS, NS-BH and BH-BH
binaries are far too small. (ii)  Exotic objects, for instance
related to dark matter candidates, are more difficult to rule out
on the basis of population synthesis models, simply because of our
ignorance of their dynamics and evolution. However, objects of
primordial origin, like primordial BHs, should reside in the halo
and in this case they would not explain a correlation between the
orientation of the bars and 
the galactic disk.
(iii) Independently of its nature, a population of sources which
produces a steady mass loss to GWs, at the level required to explain
the data, over a timescale comparable to the
age of the Galaxy, might be difficult to reconcile with the limits
from galactic dynamics discussed in this
section. Thus, an evolutionary scenario able to explain why the mass
loss is larger today than in the past would  probably be needed.

An answer to the last point might come considering sources
localized close to the galactic center, rather than distributed in the
galactic disk.  Stellar dynamics in the environment
of a supermassive black holes can have completely different timescales.
Extreme examples are given for instance by BL Lac objects, whose
luminosity can change by a factor of 100 over just a few months.
The recent discoveries of X-ray flaring 
 with a timescale of one hour from the galactic center
\cite{Baga}, and of rapid IR flaring~\cite{Ghez} point toward the
possibility of an active galactic nucleus, even if with a very low
luminosity compared to typical AGN.
In any case, for the galactic center, 
periods of enhanced activity on a timescale
of, say, millions of years, are certainly possible in principle, so it
is possible that the activity that we are seeing today is larger than
the average value
over the entire age ($t\sim 10^{10}$~yr) of the Galaxy, and then
the limits discussed above on the mass loss of the Galaxy would
disappear. 
This hypothesis has a clear-cut experimental test, to be checked on
future data: as we see from fig.~\ref{figgc}, a second peak should
appear, centered at sidereal hour 14.

\section{GW bursters\label{bursters}}

In this section we explore the possibility that all events come from a
single  source (or at most a few sources), which repeatedly emits GW
bursts. We will discuss in  sections \ref{accre}-\ref{strane} 
some possible physical
mechanisms that could produce such a ``GW burster''. First we discuss
how this hypothesis might help in the interpretation of the data.

A  motivation for this hypothesis is that it allows us to overcome
the limits on the mass loss discussed in the previous section. In
fact, if each event comes from a different source, it is very
difficult to escape the conclusion that the average 
value of the distance $r$ to
the source, which appears in \eq{1Erad}, is at least of the order of the 
distance to the galactic center, $r\simeq 8$~kpc. 
Indeed, the problem is that even within such a
distance there is not, as far as we know, a sufficiently large
population of candidate sources.

If instead we assume that the source is always the same, then the
detectors will be sensitive to the closest one, and therefore $r$ in 
\eq{1Erad} can be smaller. 
Of course, if the source that we see is ``typical'' of its population, in the
sense that its distance from us is of the order of the average
distance which could be inferred from the number of similar galactic
objects, we do not gain anything, because the factor $\sim r^2$ in energy
which we gain from the fact that the source is close  to us is
compensated by the fact that in the Galactic disk there are $\sim 1/r^2$
similar sources. The total energy radiated in GWs by the Galaxy would
then be the same.
However, when we observe just a single source,
statistical considerations do not  apply, and the source is
where it is. In this case
it therefore  makes sense to consider a lucky
situation, where we happen to be closer than expected to a source, or
where a transient phenomenon which lasts only for a short period happens
to be on. If instead each event in the EXPLORER/NAUTILUS data is
produced by a different source, the conclusion that the average
distance of the sources will be the typical one of the population should
be inescapable.

Placing the source close to us opens the possibility
that it is at work some less powerful mechanism for GW production.
For a still rather respectable distance of 
500 pc, \eq{1Erad} gives $\DE\sim 4\times 10^{-5}\msun c^2$ for each burst
while, taking $r\sim 100$~pc (which is of the order of the
distances  to the closest
known neutron stars) 
gives $\DE\sim  10^{-6}\msun c^2$.  From
population synthesis models it was estimated that the closest NS is at
about 5~pc (see ref.~\cite{ST}, pag.~12).
For a source at such a small distance 
$\DE\sim 4\times 10^{-9}\msun c^2$. (However, if the source is too
close, the fact that it is approximately in the direction of the
galactic plane would have to be explained by chance).
 Actually, the local density
of old NS is reduced by an order of magnitude by the fact that their
birth velocities are large~\cite{LLNat},  so 10~pc is probably a safer
estimate. On the other hand, the solar vicinity is enriched with young
NSs (and possibly BHs) which originate  in the Gould Belt~\cite{PTPCT}.

If the source emits
$O(200)$ bursts per year it will radiate  $10^{-2}$ solar masses per
year for  $r\simeq 500$~pc
and
$2\times 10^{-4}\msun /$yr 
for  $r\simeq 100$~pc,
and therefore it could be a transient
phenomenon which lasts for a period between a few years and a few
thousands years. 

As we have seen in sect.~\ref{sectSid}, a single source  in the
galactic plane  which emits randomly
polarized GWs has a response function  with two peaks, one close to sidereal
hour 4, and a second at a position which depends on the source
location. 
Therefore the signature for such a source would be the
emergence of a second peak when higher statistics will be 
available.
The position of the peaks can be shifted if the source emits
polarized GWs, and in this case 
the height of the second peak can be smaller than
the first.

A process which emits a
large burst of GWs should be a cataclysmic event which results
in the disruption of the source, and therefore it should be difficult
to find mechanisms that produce a GW burster. However, 
if the source is sufficiently close to us
we can explain the data with relatively small bursts, e.g.
$10^{-6}\msun c^2$ for $r\sim 100$~pc. At this level, it is possible to
imagine galactic mechanisms that produce a GW burster.
In the next  subsections we will present some possible realizations of
this idea.

\subsection{Accreting neutron stars}\label{accre}

Generally speaking, in astrophysics the presence of repeated activity
is often related to accretion onto compact
objects. An example is provided by X-ray 
bursters. These are neutron
stars which accrete matter from a companion. The magnetic field 
of the NS is
not sufficiently strong to channel the accreting matter toward the
poles, and therefore the accretion is spherically symmetric. 
Each time a layer
of about one
meter of material is accreted (which happens in a 
time  which, depending on the particular star, can be
between a few
hours and a few days) a thermonuclear flash takes place, and is
observed as an X-ray burst. These sources therefore repeatedly emit 
X-ray bursts.

\begin{figure}
\includegraphics[width=0.4\textwidth]{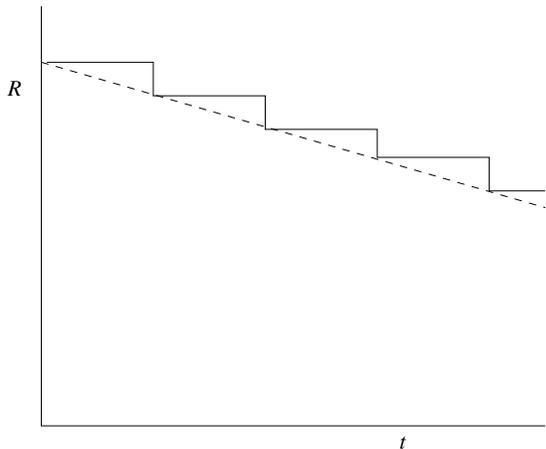}
\caption{\label{figR}A schematic drawing of 
the evolution of $R$ against time for an
  accreting NS. The dashed line is the evolution of an idealized fluid
  NS while the solid line is the evolution of a real NS with a solid crust.}
\end{figure}

Stimulated by this example, we ask whether 
a NS accreting at a steady rate can
undergo periodically some structural changes which are accompanied by the
emission of GWs. We start from the observation that
the mass-radius relation of a NS depends on the equation of
state, but it is always such that the larger is the
mass, the smaller is the radius. Thus, when a NS accretes material, its new
equilibrium radius decreases. If the NS were a fluid, a
continuous accretion of matter would produce a continuous decrease of
the radius. However, neutron stars have a solid crust, about one
km thick and with a rigidity which, in the inner part, is huge by
terrestrial standards. Therefore the radius will rather stays constant
until sufficient material has been accreted so that the crust can be
broken, and the
evolution of the radius will
rather be a sequence of jumps,  as shown schematically in fig.~\ref{figR}.

In each jump a certain amount of energy is released.
Two questions are therefore important for our purposes: how much
energy is released per jump, and whether this energy can be radiated
away in GWs. 

Neutron star perturbations can be described in terms of quasi normal modes, 
which can be excited by many possible mechanism: accreting material, 
crust breaking, starquakes, onset of phase transitions, coalescence, etc. 
These perturbations
can excite in particular the neutron star quasi normal modes described
by spherical harmonics with $l=2$, which
dominates the emission of GWs. 

Furthermore, if the NS is rotating, even the radial oscillation 
will induce a time varying quadrupole moment, and in this case
energy will be liberated in GWs, with a
damping timescale~\cite{Chau}
\be\label{tau}
\tau_{\rm GW} \simeq 9 {\rm ms}\, \frac{M}{\msun} 
\(\frac{P_{\rm rot}}{1\,\rm ms}\)^4
\(\frac{14\,\rm km}{R}\)^2\, ,
\ee
where $P_{\rm rot}$ is the rotational period. 
We have taken 14~km as a reference value for $R$, since this is the
typical value of the radius for a fast rotating NS in a large range of
masses, $0.3<M<1.2\msun$,
see e.g. sect.~7.1 of ref.~\cite{Gle}. 

For a NS spinning with $P_{\rm rot}$ of the
order of a few ms, $\tau_{\rm GW}$ is smaller than
the damping time due to viscosity, and the energy liberated will be
radiated away mainly in GWs (see also refs.~\cite{CD,MVP}).

The next question is how much energy is liberated in each jump. The
source of energy is the potential energy $U$ of the NS. In order of
magnitude, $U\simeq (3/5) GM^2/R$ and, if the crust breaks when a mass
$\D M$ has been accreted, the energy liberated is
\be
\frac{\D U}{U}\sim 2\frac{\D M}{M}-\frac{\D R}{R}\, .
\ee
The two contributions go in the same direction since a positive $\D M$
induces a negative $\D R$. However, for typical equations of states,
especially for rotating neutron stars, $|\D R|/R$ is smaller than $\D
M/M$ (see e.g. fig.~7.1 of ref.~\cite{Gle}), so we just set
$|\D U/U|\sim 2 \D M/M$. As an order of magnitude
\be
|U|\sim \frac{3}{5}\frac{GM^2}{R}=
\frac{3}{10}\(\frac{2GM/c^2}{R}\) \, Mc^2\sim 0.1 Mc^2\, 
\ee
for a typical NS. Therefore 
the energy that is released when the crust of a NS collapses under the weight
of a mass $\D M$ can be roughly estimated as 
$E_{\rm rad}\sim |\D U|\sim 0.2\D M c^2$. To estimate the value of $\D M$
which induces the collapse of the crust 
we consider the pressure $P$ exerted by the mass
$\D M$,
\be
P= \frac{GM\D M}{R^2}\, \frac{1}{4\pi R^2} =\frac{GM\D M}{4\pi R^4}
\ee
and we equate it
to the maximum shear stress $\s_{\rm max}$ that can be sustained by
the NS crust. The latter
 has been investigated  in the context of
pulsar glitches \cite{Rud,BP,RudI,RudII,ST,SW};  the  maximum shear
stress can be written as~\cite{RudI}
\be
\s_{\rm max} = \frac{L}{R}\mu\,\th_{\rm max}\, ,
\ee 
where $L\simeq 1$~km is the thickness of the crust, $R$ is the NS radius,
$\mu$ is the shear modulus and
$\th_{\rm max}$ is the maximum strain angle  that the crust can sustain
without breaking. In the lower part of a 1~km thick NS
crust, it is estimated that
$\mu\sim 10^{30}$~cgs, while in most of the crust region it is 
of order $2\times 10^{29}$~cgs~\cite{RudI}. Concerning $\th_{\rm max}$,
experience with very strong terrestrial
materials gives
$\th_{\rm max}\sim 10^{-2}$ (only after hardening and at low
temperatures). However, the actual value of $\th_{\rm max}$ is
probably lowered by dislocations, and could be of order
$\th_{\rm max}\sim 10^{-5}$ to $10^{-3}$ \cite{SW}. Therefore, taking
$L/R\simeq 0.1$, we get
\be
\D M\sim\, 4\times 10^{-9}\msun\, \( \frac{R}{14\,\rm km}\)^4
\(\frac{\msun}{M}\)\(\frac{\th_{\rm max}}{10^{-3}}\)\, .
\ee
and correspondingly the maximum energy that can be liberated in a
single starquake is
\be\label{Equake}
\D E_{\rm rad}\sim
\,  8\times  10^{-10}\msun c^2\, \( \frac{R}{14\,\rm km}\)^4
\(\frac{\msun}{M}\)\(\frac{\th_{\rm max}}{10^{-3}}\)\, .
\ee
Taking the most favorable values $M\sim 0.5\msun$
and $\th_{\rm max}\sim 10^{-2}$ we can arrive   to
$\D E_{\rm rad}\sim 10^{-8}\msun c^2 $. 
However, if
we use the more plausible estimate 
$\th_{\rm max}\sim 10^{-5}$ to $10^{-3}$,  as well as $M\sim 1.4\msun$
(which in turn gives a smaller value of $R$,
$R\sim 10-12$ km~\cite{Gle}), we obtain  much smaller values,
$\DE\sim 10^{-12}$ to $10^{-10} \msun c^2$.

As we discussed 
above, even placing the source extremely close to us, say
$r\sim 5$~pc (corresponding to the estimated distance to the closest
NS), we still need a mechanism which radiates at least
$4\times 10^{-9}\msun c^2$.  So
it seems that, even with such an extreme choice for $r$,
it is not easy to explain the events seen by EXPLORER/NAUTILUS by
the breaking of a neutron star crust.
Furthermore,  a very close accreting NS should have
been seen as an X-ray source.

Another possibility, however, is that the
starquake due to the crust breaking, rather than being the
source of the GWs, might be the trigger for some more important structural
changes inside the NS. In particular, the core of a NS can perform a phase
transition from a hadronic to a deconfined quark-gluon phase. In the
literature ~\cite{CD,MVP} it has been
considered the possibility that an accreting NS acquires a sufficient
mass to perform completely the transition from a hadronic core to a
quark-gluon core. The energy gained in the transition, for a NS with
a mass $M\sim 1.5\msun$, is of order $0.15\msun c^2$~\cite{MVP}, and it has
been observed that this large energy can excite quasi normal modes 
and  be liberated in a  GW bursts  (with 
high efficiency if the NS rotates sufficiently fast). 
Particularly interesting appear the discontinuity $g$-modes ~\cite{Valeria}, 
having typical frequency 
in the range 0.5-1.4 kHz and constituting an
unique probe for density
discontinuities, like the 
ones induced by phase transitions.
One can imagine  that
the phase transition does not take place suddenly and completely in
the whole core. Rather, each time the critical mass $\D M$ is
accreted, a starquake takes place and  transforms successive layers of the NS
core from the hadronic to the deconfined phase. In this case the
$0.15$ solar masses will not be released in a single, very large
bursts, but rather in a  series of bursts. 
 One can imagine variants of this
scenario in which the phase transition involves strange quark
matter, hybrid quark stars~\cite{Gle},
or a phase of color crystallization~\cite{ABR}.
In all cases, unfortunately, a  computation of the energy liberated by the
phase transition in a NS layer 
as a consequence of the starquake appears a rather
difficult task, so this scenario remains quite speculative.

\subsection{Soft Gamma Repeaters}

Soft Gamma Repeaters (SGRs) are X-ray sources with a persistent
luminosity of order $10^{35}-10^{36}$ erg/s, that occasionally emit
huge bursts of soft $\g$-rays, with a power up to $10^{42}$ erg/s, for
a duration of order $0.1$~s. There are three known SGRs in
the galactic plane (SGR1900+14, SGR1806-20, SGR1627-41) and 
one in the direction of the galactic center, 
SGR1801-23. Furthermore, SGR0525-66 is located in the Large Magellanic Cloud.

These  objects  are understood as magnetars
\cite{DT,TD1,TD2,TD3}, i.e., as magnetically powered
neutron stars with huge magnetic
fields of order $10^{14}-10^{15}$ gauss. 
The hypothesis that SGRs are
neutron stars is supported  by the hardness and luminosity of the
bursts,  by the
periodic modulation of the soft tail and
by the fact that for at least three of them have 
been found associations with young ($t\sim 10^4$ yr) supernova
remnants.
Experimental 
evidence for these huge values of the magnetic field comes
from the observed spin-down rate~\cite{Kou}.
The magnetar model connects SGRs to
another  class of enigmatic objects, anomalous x-ray pulsars (AXPs),  
which brings
the number of known candidates to about twelve~\cite{Dunhome}.

Magnetic field lines
in magnetars drift through the liquid interior of the NS, stressing
the crust from below and generating strong shear strains. For magnetic fields
stronger than about $10^{14}$ gauss, these stresses are so large that
they cause the breaking of the  1~km thick NS crust, and
this elastic energy is suddenly released  in a large
starquake, which generates a burst of soft gamma rays.

The statistical behavior of SGRs is
strikingly similar to that of earthquakes: they obey the
power low energy distribution of Gutenberg and Richter, and they have the same
waiting time distribution~\cite{Cheng}. 
The bursts arrive in bunches, when the crust is yielding to the
magnetic stresses. For instance SGR1900+14, after decades of quiescence,
emitted over 50 detected burst during the last week of May 1998, and
continued to burst into early June. It emitted a giant flare  on
Aug.~27, 1998 (see below) and overall during nine months in 1998 it
emitted over 1000 detected bursts.
The source SGR1627-41 instead suddenly showed up
emitting about 100 bursts  in June-July 1998~\cite{Dunhome}.

Occasionally, truly giant flares
have been detected. One is
the March 5, 1979  event from SGR0525-66 in the LMC,
which fueled interest in SGR astronomy; this source emitted overall 16  bursts
until May 1983, when the activity ceased. No more bursts have been
detected from this source since.
The other giant flare is the event on
Aug.~27, 1998  from SGR1900+14~\cite{Hur}. Another very 
bright bursts was emitted by this source on Apr.~18, 2001, and several
more commons bursts where detected in the following weeks,  including
another large burst
on Apr.~28, 2001 (see \cite{Wood} and references therein). 
Giant flares liberate more than $10^{44}$ erg
(i.e. $10^{-10}\msun c^2$) in gamma rays, and have a longer duration,
of order 100~s. In the magnetar model they are believed to be
produced by a global large-scale rearrangement of the magnetic field, while
smaller bursts are produced by local ``crustquakes'' of the NS.

The emission of GWs from  magnetars has been studied
in refs.~\cite{FP2,Ioka}. The breaking of the crust 
produces shear waves~\cite{Blaes}
with a period of the order of the ms, that excite non-radial
oscillation modes of the NS, damped by the production of GWs with a
frequency of the order of the kHz.
For ``normal'' bursts, ref.~\cite{FP2} estimates that the total
elastic energy that can be released is of order 
$10^{45}\, {\rm erg}\simeq 5.6\times 10^{-10}\msun c^2$, in very good
agreement with our \eq{Equake} (since, independently of the agent which
causes the  crustquake, either accretion of magnetic fields,
this is the maximum elastic energy that can be released by the
breaking of the  crust).
Ref.~\cite{Ioka} performs a
detailed analysis of the equilibrium configuration of a magnetic polytrope 
and finds that, in giant flares,  
the total energy  released by the magnetic field
rearrangement can be $E_{\rm tot}> 10^{47}
\, {\rm erg}\simeq 5.6\times 10^{-8}\msun c^2$, and possibly even as
high as  $E_{\rm tot}> 10^{49}
\, {\rm erg}\simeq 5.6\times 10^{-6}\msun c^2$, for extreme values of
the parameters of the model. This energy could by radiated in GWs with
high efficency.

The distances to observed SGR are highly
uncertain, but they should all be at $r>5$~kpc. In particular
the distance to 
SGR1900+14 is estimated at $r\sim 5-6$~kpc. 
Comparing with \eq{1Erad} we see that, if this is correct, 
the energy requirements are not met since, for sources at 5~kpc, we
need a process which emits more than
$10^{-3}\msun c^2$ in GWs.

Nevertheless, magnetars could  become very interesting candidates for
GW production
if it were possible to imagine that one or more of these sources are
very close to us. Of course, since no very close source has been
detected, we must provide a mechanism which forbids the
observation of
electromagnetic emission, while leaving the GW emission. One generic 
possibility that 
comes to mind is that the $\g$-ray emission might be beamed. This cannot
be excluded a priori in the case of the ``normal'' flares, where the $\g$-rays
originates from localized cracks in the NS crust.

A second possibility is related to the fact that 
a critical value of the magnetic field 
$B\gsim 3\times 10^{14}$ gauss
is required to
suppress the electron cross section on one photon polarization state
below the Thomson value. This decrease in the scattering opacity
allows the photon luminosities to reach the super-Eddington values
$L\sim 10^4L_{\rm Edd}$ observed in SGR~\cite{Pac,TD2}. 
Neutron stars with a magnetic field
below this critical value would be much less bright
electromagnetically, but might still have sufficient tectonic activity
to produce a significant amount of GWs.

A distinctive and testable feature of the  hypothesis that the
EXPLORER/NAUTILUS signals come from SGRs is a temporal
clustering of the events, since each source has long periods of
quiescence (years or decades) until it suddenly enters in a phase of
intense activity, with bursts arriving in bounces on the timescale of
hours to 1-2 months. From this point of view, it is interesting
to observe that, out of the eight events at the sidereal hour peak in
2001, two came at the distance of one hour from each other 
(see Table~3 of ref.~\cite{ROG01}). If these events correspond to real
GW signals, it is very unlikely that they came from two
different sources, and they would rather suggest a single source with
repeated activity.

Magnetars are believed to stay 
in their SGR phase in the first $\sim 10^4-10^5$ yr
of their life. Then, when the star cools below  a threshold, the
dissipation of magnetic activity ceases. This is suggested by the
theory~\cite{TD3} and also fits well with the estimated
ages of the supernova remnants
which are believed to be associated with SGRs. When the dissipation of
magnetic energy ceases the NS enters the so-called ``dead magnetar'' phase,
but the theory suggests that these stars remain strongly magnetized. 
It is suggested that a large fraction of all NS, say one-half, have indeed
been active magnetars~\cite{Dunhome}. In this case, the number of NS highly
magnetized but  presently 
magnetically inactive  would be of the same order as ``normal'' NS. Recalling
that the distance to the closest NS is estimated 
to be of order
10~pc, the same order of magnitude estimate would hold for dead
magnetars. 
If  these objects   maintained a  starquake
activity, possibly related to a residual slow diffusion of the
interior magnetic field, they would certainly be very interesting
candidates as GW sources.

\subsection{Strange quark stars}\label{strane}

Another possible realization of a GW burster is provided by the $r$-mode
instability in strange stars (see chapter~12 of ref.~\cite{Gle} for an
introduction to strange stars).
As discussed in ref.~\cite{AJK}, in
stars made of strange quark matter the $r$-mode instability has a
dynamics quite distinct from the neutron star case. In particular, in
an accreting strange star the evolution of the GW amplitude generated
by the $r$-mode during its first year of evolution consists of a
repeated series of bursts on a timescales from hours to months (see 
in particular fig.~4 of ref.~\cite{AJK}). 
The GW amplitude 
has been estimated  to be (eq.~(32) of ref.~\cite{AJK}, with
$\tilde{J}=1.635\times 10^{-2}$ for an $n=1$ polytrope)
\be\label{hstrange}
h\sim 2\times 10^{-20}\a\, \(\frac{1\,\,{\rm kpc}}{r}\)
\(\frac{M}{1.4\msun}\)  \(\frac{1\,{\rm ms}}{P_{\rm rot}}\)^3 
\(\frac{R}{10\,{\rm km}}\)^3 ,
\ee
where $P_{\rm rot}$ is the rotation period and $\a$ is the $r$-mode amplitude. 
During the first year of evolution of a young strange star the
parameter $\a$ performs 
large oscillations from very low values,
$\a\sim 10^{-15}$, up to values of order one.
These rapid variations therefore result in a series of GW bursts. The
amplitude in \eq{hstrange}, for a distance $r\sim 1$~kpc, is still too
small compared to the values $h\sim 2.5\times 10^{-18}$ which
corresponds to the NAUTILUS/EXPLORER events when the energy deposed in
the bar is $E_s=100$~mK, see \eq{h2518}. 
Still, as stressed in ref.~\cite{AJK}, the
calculation of the GW signal in this early phase of the evolution
of the strange star  is
very model dependent and there are large theoretical uncertainties. 

The main problem with this source, however, is that within the model
discussed in ref.~\cite{AJK}
this kind of activity can take place  only for very young quark stars,
i.e., in their first year  of evolution. Since
for the application to the EXPLORER/NAUTILUS data we  need a galactic,
and even relatively close source,  this
mechanism cannot explain the EXPLORER/NAUTILUS result, unless there is
a way to rise significantly the duration of the activity,
as is the case for SGR.

\subsection{Search for periodicities}

If the events are all due to different sources, there will be no time
correlation between  events. However, if all the events are due
to a single (or a few) GW burster, it makes sense to explore the
possibility that there is a periodicity in the arrival times of the
data.
An almost periodic behavior could be generated by many
different mechanisms. For instance, X-ray bursters show a regular
behavior since they accrete at a uniform rate (often
regulated by the fact that it is at the Eddington limit), and go off
each time that a given critical mass has been accreted.
So, if detected, such a periodicity would 
be really a ``smoking gun'', showing
that the events come from a single source. We have therefore looked if
a fit of the form $t_n=t_0+n P$ can reproduce the arrival time of the
coincidences reported in ref.~\cite{ROG01}, for sidereal hours 3 and 4.
The analysis is complicated by
the fact that,
even if such a periodicity existed,  most of the
values $t_n$ would correspond to times where either the two detectors
were not simultaneously on, or they were not well oriented with
respect to the source. Furthermore, in the two bins corresponding to
sidereal hours 3 and 4,
two events are expected to be background, and therefore introduce spurious
correlations. 
Also,   one should not necessarily look for an exact periodicity
but rather for some regular activity, i.e., something of the form 
$t_n=t_0+n P +\D t_n$, with $\D t_n\ll P$, since in general the
mechanism that produces the bursts will not be an exact clock, but
rather will be related to some physical process, like accretion, which
proceeds more or less at a steady rate.

Defining $f(\o )=\sum_n\exp\{ -i\o t_n\}$, where the $t_n$ are the
arrival times of the events (and restricting to the 8 
events at sidereal hours
3 and 4), for a periodic behavior with period $P$ the function
$|f(\o )|$ should have a peak at $\o =2\pi /P$.  Of course, a plot of 
$|f(\o )|$ will show very many oscillations due to noise,
and the point is whether there is
a peak which emerges clearly above this noise.  Within the statistics
avaliable, we find that this is not the case. 

To understand more quantitatively what this negative result means, we
consider one of the highest peak of $|f(\o )|$ (which corresponds to a
period $P\simeq 4.473$ days), and we verify that with this value of $P$
we are able to reproduce the arrival times of 6 out of the 8 events
with a precision of $\pm 3$~hours. We do not assign any positive
significance to this result, since $|f(\o )|$ displays  many other
peaks of approximately the same intensity. Rather, we can use this
result to exclude the existence, in the data, of
very precise periodicities, i.e., writing
$t_n=t_0+n P +\D t_n$, we can say that the data
exclude a fit of this form, with $|\D t_n|\lsim 1$~hr. Further
statistics will be needed to examine the existence of approximate 
periodicities with $|\D t_n|\gsim 1$~hr.

This negative result still gives
useful informations, because it allows us to exclude
mechanisms that would produces an extremely regular
series of bursts.

\subsection{Cosmic strings}\label{sectCS}

A completely different physical
scenario is provided by loops of cosmic
strings. Cosmic strings are topological defects which appears in
various grand unified theories and which might be nucleated 
at a symmetry breaking phase transition in the
early Universe (for review, see \cite{Vilbook}).
It has long been known that they can be a source of
stochastic background of GWs of cosmological origin~\cite{Vil}.
More recently, stimulated by a suggestion of ref.~\cite{BHV},
it has been shown in refs.~\cite{DV1,DV2} that cusps in cosmic string loops
can emit GW bursts whose amplitude, depending on the parameters of the
model (in particular,
the string tension and the number of cusps per loops) can be
interesting for GW detection.

However, in our case an interpretation in terms of cosmic strings
suffers of two main problems. The first is that, even with the most
optimistic values of the parameters, the amplitude for the GW burst
in the kHz region does not exceed $h=10^{-21}$, see fig.~1 of
ref.~\cite{DV2}. While this can be barely detectable with the target
sensitivities 
of the LIGO and VIRGO interferometers, it is much smaller than the
amplitude corresponding to the EXPLORER/NAUTILUS data which, for 
a deposited energy $E_s=100$~mK, is rather  $h\simeq
2.5\times 10^{-18}$.

Second, cosmic string loops are cosmological objects, and therefore 
there is no reason for obtaining
a correlation between  the orientation of the bars and the 
the galactic plane.

\subsection{Summary}

In this section we have examined the possibility that all the $O(200)$ 
events per year inferred from the EXPLORER/NAUTILUS observations come
from a single, or at most a few, GW bursters. We have examined different
possible realizations of this idea and, in general, we find that
the idea is viable only if the source is very close (unless the GW energy
liberated in each burst is much larger than the estimates that we 
presented). To put
the source very close, we need a mechanism which shuts off the other form
of radiation, like X-rays or $\g$-rays, otherwise such close sources would
be very bright electromagnetically.

Accretion onto neutron stars does not seem to offer this
possibility. However, magnetically powered neutron stars (magnetars) have a
dynamics which depends critically on the value of the magnetic field
and might offer a broader range of possibilities. Further
investigation in this direction is certainly worthwhile.

Magnetars  also have the rather special property that, after long
periods of relatively quite life, they suddenly become active and emit
signals  arriving in bounces on a timescale of
hours to months.  Certainly this will be an important point to
be tested on future data from long runs.

\section{Acoustic detection of massive particles}\label{sectAcou}

Beside the issue of the statistical significance of the results of the
2001 run, which will be hopefully settled in the near future, it is
also crucial to understand whether the excitations of the bars
could be due to some phenomenon unrelated to GWs.
One important source of concern is the fact that
particles crossing a bar can lose energy by recoil against the
nuclei of the material. This will produce a  warming up and a local
thermal expansion, which in turns originates mechanical vibrations in
the bar. The vibrational energy $E_s$ deposed in the fundamental mode of
a cylindrical bar by this mechanism has been studied by several
authors \cite{BH,GST,AC,BRL,DRG,AP,LB} 
and is given by the formula
\bees
E_s &=& \frac{4\g_G^2}{9\pi\rho Lv^2}\(\frac{dE}{dx}\)^2
\times\nn\\
&\times &\[
\sin\(\frac{\pi z_0}{L}\)\sin\(\frac{\pi l_0\cos\th}{2L}\)
\frac{L}{\pi R\cos\th}\]^2\label{acoustic}\, ,
\ees
where $L$ is the bar length, $R$ is the bar radius, $l_0$ the length
of the particle's track inside the bar, $z_0$ the distance of the
track midpoint from one end of the bar, $\th$ the angle between the
particle track and the axis of the bar, $dE/dx$ the energy loss per
unit length of the
particle in the bar, $\rho$ the bar density, $v$ the sound velocity of
the material and $\g_G$ the Gr\"uneisen coefficient of the material,
which depends on the ratio of the thermal expansion coefficient to the
specific heat ($\g_G\simeq 1.6$ for Aluminum at low temperatures).

This formula has been well verified experimentally for cosmic rays,
when the resonant bar is
in a non-superconducting state (which is the mode in which the bar is
operated in the 2001 run that we are discussing), 
while it appears to underestimate
$E_s$ when the bar is in the superconductive state~\cite{Ast}.

Cosmic rays are detected by two  detectors placed above and
below each resonant bar, and the corresponding events are
eliminated from the list of GW candidate events. However a massive, electrically neutral, long-lived
particle
could avoid the cosmic ray veto and  depose energy in
the fundamental vibrational mode of
the  bar, according to \eq{acoustic}. In this section we therefore
explore whether it is possible for  massive particles
to produce a signal compatible with the EXPLORER/NAUTILUS
observations.

If the existence of a correlation between the energies deposed in the two
bars will be confirmed by further data, it will be extremely difficult to
see  how such a result could be originated by
particles whizzing around,
such that occasionally one particle hits one bar while a second
particle hits the other bar.  Observe in particular, from table~3 of
ref.~\cite{ROG01}, that for each of the eight events at sidereal hours 3 and 4,
the energy $E_{\rm Expl}$ deposed in EXPLORER and the
energy $E_{\rm Nau}$ deposed in NAUTILUS always differ by less than 20\%
and in some cases by about 5\%. From one event to the other, instead,
the energy can change by as much as a factor of 4.
Similarly,  it is difficult to see how a sidereal time
modulation could emerge from the interaction with a random flux of
particles. 

A dependence on sidereal time could instead be understood
if the two bars are detecting 
particles which are ejected in  astrophysical explosions.
Even in this case, it is not easy to explain the existence of a
strong correlation in energy. One should make the
assumption that the particles are produced 
with an energy spectrum which is strongly peaked, so that 
the typical spread in energy is small, with 
$\D E/E\sim 5-10\%$. The position of the peak in the energy spectrum
might depend on the conditions in the star, so it could  change
from event to event (as it is observed in the data), 
still maintaining a correlation between the energies
detected in the two bars. Such a scenario 
does not seem very
plausible; nevertheless, let us assume it for the moment and see if we
can find more quantitative arguments in order to
rule it out.

First of all observe that in this scenario
the particles come from astronomical distances and  are
massive, so any spread $\D v$ in their velocities will produce a
corresponding spread $\D t$ in the arrival times. Since the bursts seen
by the bars are concentrated in an interval $\D t<0.1$~s, the particles
must be highly relativistic, so that their spread $\D v$ is
sufficiently  small. In terms of $\g =(1-v^2/c^2)^{-1/2} $  we have, 
for $v$  close to one,
\be
\frac{\D v}{v}\simeq \frac{1}{\g^2}\(\frac{\D \g}{\g}\)\, .
\ee
Since $\D \g/\g =\D E/E$ and we are assuming
that $\D E/E\,\lsim 1$ in order to account for the energy correlation, 
we have $\D v/v\sim 1/\g^2$ and
therefore $\D t/t\sim 1/\g^2$, where $\D t$ is the duration of the
bursts, and $t\simeq r/c$ is the time taken by the particle to arrive
from a distance $r$. Requiring $\D t<0.1$~s gives
\be\label{g106}
\g > 10^6\, \(\frac{r}{1\,\rm kpc}\)^{1/2}\, .
\ee
Therefore, for  astronomical distances, we must have highly relativistic
particles. 

\begin{figure}
\includegraphics[width=0.4\textwidth]{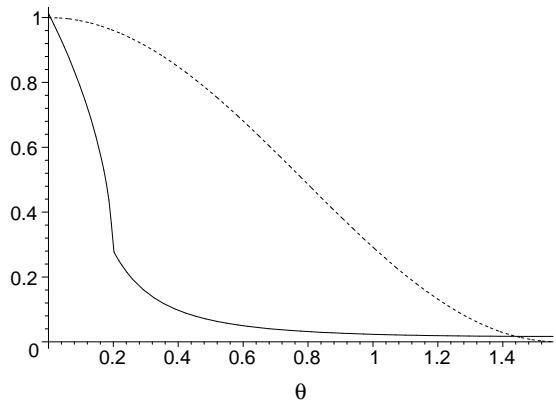}
\caption{\label{figDMth} The sensitivity curve of the bar 
for the acustic detection of massive particle, as a function of the
incoming angle $\th$ (solid line). To appreciate the fact that the
response
function is quite narrowly peaked, we also plot for comparison the function
$\cos^2\th$ (dashed line). The cusp at
$\th =2R/L\simeq 0.2$ reflects the fact that this is the maximum angle
for which the particles traverse the bar longitudinally.
We averaged over all possible impact points on the  surface of the bar.
}
\end{figure}

Observe that
\eq{acoustic} has a strong
angular dependence which is due to the fact that a particle which
crosses the bar along its longitudinal axis has a longer path inside
the material, and therefore a larger energy loss, compared to a
particle which goes through the bar transversally.  
When a particle enters at an angle $\th =0$ (i.e. in the longitudinal
direction)  we have $l_0=L, z_0=L/2$, and the term
$\[ \ldots \]^2$ in \eq{acoustic} becomes $L^2/(\pi R)^2$. 
On the other hand, for a particle entering perpendicularly to the bar
we have $\cos\th\ra 0$ and
$l_0=2R$ so, in \eq{acoustic}, $\[ \ldots \]^2\ra  \sin^2(\pi
z_0/L)\ra 1/2$ (averaging the sin squared over
$z_0$). Therefore the energy deposed by particles impinging on the bar 
longitudinally is larger than for particles impinging transversally by
a factor $2L^2/(\pi R)^2$, which is of order 20 with 
the values $L\simeq 3$~m, $R\simeq 0.3$~m of the bars. 
The dependence on $\th$ of
the response function, shown
in fig~\ref{figDMth}, is quite peaked, since the maximum angle for
which the particles traverse the bar in the longitudinal direction is 
$\th=2R/L\simeq 0.2$. Therefore, even as an acoustic
detector of particles, the bar is sensitive to the orientation with
respect to the source.

It is useful to separate the discussion into two
parts, corresponding to particles traversing the bar along the
longitudinal or the transverse direction.

\subsection{Longitudinal  trajectories}

Since the excess of coincidences is seen when
the bar is orthogonal to the galactic plane, particles traversing the
bar longitudinally would come from the directions of the galactic poles.
So we should  ask what
lies in  these directions and an answer is that,
very close to the North Galactic Pole, there is the Virgo cluster of
galaxies.
This result is quite interesting because, with the 2500
galaxies present in the Virgo cluster (which include giant galaxies) 
a rate of order 200 events per year could
be explained by explosions taking place at a rate of a few events per
century per galaxy, quite compatible, for instance, with the rate of
supernova explosions.

\begin{figure}
\includegraphics[width=0.3\textwidth, angle=270]{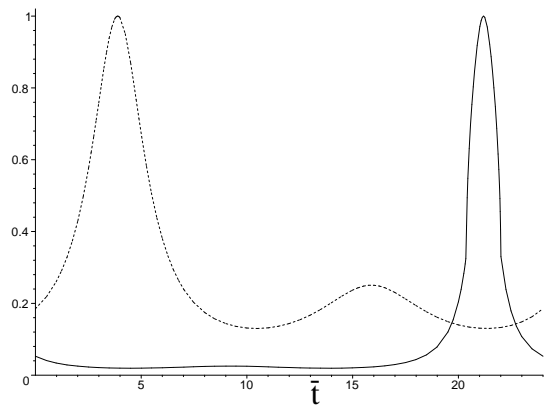}
\caption{\label{figDMVirgo} The response function of the bar 
for the acustic detection of massive particle, for a source located
  at the center of the Virgo cluster (dashed line), and 
for a source in the galactic
  center (solid line), as a function of sidereal time $\bt$.
}
\end{figure}

In fig.~\ref{figDMVirgo} we show the
sensitivity curve of the bar as an acoustic particle detector, 
i.e., we show the energy deposed by a particle in the bar as
a function of sidereal time, for a fixed energy of the incoming
particle and a given source location.  The sensitivity 
to
sources located in the Virgo cluster (dashed line) is peaked at 
sidereal hour 3.9 (using as always the average sideral time between
EXPLORER and NAUTILUS). For comparison, we also show
the sensitivity to sources in the galactic center (solid line), which has a
maximum at sidereal time 21.
In both cases, the peaks are quite narrow. The sensitivity curves have
been normalized so that the value at the peak is one.

Inserting the numerical values in \eq{acoustic} we find that a signal
$E_s\simeq 100$~mK can be obtained from a particle which crosses the
bar longitudinally, with an energy loss in Aluminium
\be
\frac{dE}{dx}\simeq 20\, \frac{\rm GeV}{\rm cm}
\ee
or, in a generic material,
\be\label{stopping}
\frac{1}{\rho}\frac{dE}{dx}\simeq 7\, \frac{{\rm GeV}\, {\rm cm}^2}{\rm
  gr}\, .
\ee
Traversing the bar  longitudinally ($L=3$~m)
the particle will then lose approximately
$6\times 10^3$~GeV,
and a similar amount will be lost  traversing
the atmosphere (since $\rho_{\rm Al}\times (3\, {\rm m})\sim \rho_{\rm Air}
\times (10\, {\rm km})$); therefore the particle
must have had an initial
energy of at least $E\sim 10^4$~GeV.
This energy arrives on a surface $\pi R^2\sim
0.3\, {\rm m}^2$.  
The acceptance  of the detector for particles traversing in the
longitudinal direction is 
$A_L=\pi R^2 \d\O$ where  $\d\O$ is the solid angle subtended by
the particles.  For particles traversing the bar in the 
longitudinal direction  
the solid angle is quite small,
$\d\O\simeq  (2R/L)^2\,{\rm sr} $ 
and  
$A_L\simeq 1.1\times 10^{-2}\, {\rm m^2}\, {\rm sr}$. 
Then
the total energy radiated in the burst must have been
\be\label{EV}
E_{\rm tot}\simeq E\,\frac{4\pi r^2}{A_L}\simeq 
 10^{-11}\, \( \frac{E}{10^4\,\rm GeV} \)
\( \frac{r}{1\,\rm kpc} \)^2 \, \msun c^2\, .
\ee
In order to obtain an astrophysically realistic value we require that 
$E_{\rm tot}<10^{-2}\msun c^2$. This gives
\be\label{E102}
1\,\lsim \, \frac{E}{10^4\,\rm GeV} \,\lsim  \,
\( \frac{32\,\rm Mpc}{r}\)^2\, ,
\ee
where the lower bound comes from the fact that and initial particle energy
$E\gsim {10^4\,\rm GeV}$ is needed in order to depose $100$~mK 
in the bar. \Eq{E102} also implies a maximum value for the distance to the
source, 
$r\lsim 32$~Mpc; this distance
include the Virgo cluster, which is approximately
at $r\sim 16$~Mpc.
Writing $E=\g m$ and using \eq{g106} we also find 
from \eq{E102} a limit on the
particle mass,
\be\label{m107}
m <  10^7 
\(\frac{1\, \rm kpc}{r}\)^{5/2}\, {\rm GeV}\, .
\ee
For a source in the Virgo 
cluster at  $r=16$~Mpc this gives $m<0.3$~MeV.

Our final information on these hypothetical
particles is on their flux. The event
rate, as we have seen, is $O(200)$ events per year.
Using the acceptance $A_L$ found above
the flux should be
\be
F\sim \frac{200}{\rm yr\,\times 1.1 \times 10^{-2}\, {\rm m^2}\, {\rm sr}}
\simeq   6\times 10^{-8}\, {\rm cm}^{-2} {\rm s}^{-1} {\rm sr}^{-1}\, .
\ee
Summarizing the above discussion, one might formulate the 
hypothesis that
the EXPLORER/NAUTILUS data could  be
due to electrically neutral
massive particles coming from  explosions in the Virgo
cluster. In this case  the sidereal time dependence could be
explained, and one could as well find a sufficient number of sources to
justify the observed rate. The candidate particle should have 
$\g >10^8$ (from
\eq{g106} with $r=16$~Mpc), a mass $m<0.3$~MeV,  an energy
$10^4\,{\rm GeV}<E<4\times 10^4$~GeV,
a stopping power 
$(1/\rho) dE/dx= 7 {\rm GeV} {\rm cm}^2/{\rm gr}$ and a flux
$F\simeq  6\times 10^{-8}\, {\rm cm}^{-2} {\rm s}^{-1} {\rm sr}^{-1}$.

The existence of a particle with these properties
is however excluded.
WIMPS are obviously out of question: with their
typical energy loss on nucleons of order 100~keV
and  weak cross sections, WIMPS have
$(1/\rho) dE/dx\sim 10^{-27} {\rm GeV} {\rm cm}^2/{\rm gr}$.
More in general, if a particle 
with a stopping power 
$(1/\rho) dE/dx= 7 {\rm GeV} {\rm cm}^2/{\rm gr}$ 
and a flux 
$F\simeq  6\times 10^{-8}\, {\rm cm}^{-2} {\rm s}^{-1} {\rm sr}^{-1}$
existed, it would have
been detected by experiments on cosmic rays at
altitude, which for these stopping powers 
are sensitive to fluxes many orders of magnitudes
below our required value.
The precise value of the experimental upper
limit on the flux  depends on the specific properties
of the particle which
is being searched. However, to get an idea of the order of magnitudes,
we may observe that cosmic ray experiments have put limits on the
fluxes of hypothetical objects like
nuclearites (bound states of $u,d,s$ quarks surrounded by an electron
cloud) and Q-balls. These exotic objects are extremely massive and are
therefore excluded as our candidate particle. However their
energy losses in matter, which is the main parameter for detection, 
are comparable to the value that we need.
The upper limit on the flux of these particles are of
order $ 10^{-14}\, {\rm cm}^{-2} {\rm s}^{-1} {\rm sr}^{-1}$ from
altitude experiments and $ 10^{-16}\, {\rm cm}^{-2} {\rm s}^{-1} {\rm sr}^{-1}$
from underground detectors like MACRO (see ref.~\cite{GS} for review).

Furthermore, a strongly interacting particle with a mass below the MeV would
spoil primordial nucleosynthesis.

\subsection{Transverse  trajectories}

Recall from the discussion of 
sect.~\ref{sectunp} that the sensitivity curve determines the
distribution of the number of events versus sidereal time only when
the typical SNR is low. In the opposite limit  
where almost all events
are above threshold, an event will be detected independently of its
arrival direction. In this case the bar will detect more events when the
particle flux is transverse to the bar axis simply because
the geometric acceptance for transverse particles is much larger than
for longitudinal particles. In fact, the 
geometric cross section
is now $S=2RL\simeq 1.8\, {\rm m}^2$ 
and the solid angle is of order
$4\pi$ (more precise estimates are not necessary for our order of
magnitude calculation).
Therefore the acceptance for  particles arriving transversally is
$A_T\simeq 4\pi (2RL)\simeq 22.6\, {\rm m}^2\,{\rm sr}$, to be
compared with the acceptance for particles arriving longitudinally,
$A_L\simeq 1.1\times 10^{-2}\, {\rm m}^2\,{\rm sr}$, as found in the
previous subsection.

In the case of sources
with high SNR in the galactic plane  the dependence of
the number of events on $\th$ would then be proportional to $\sin\th$,
because $L\sin\th$ is the projection of the bar length in the
direction orthogonal to the galactic plane. 
In a two-detector correlation the number of coincidences will be
proportional to $\sin^2\th$; this function has
a maximum around sidereal hour 4, so the location of the peak 
can be reproduced, but the peak in this case is rather broad, 
so this scenario is is not favored
by the present data, which rather suggest the presence
of  a relatively narrow peak,
concentrated in two hours. 

The analysis of the possible particle physics candidates proceeds
similarly to the previous sections, with two main differences: first,
because of the angular dependence of \eq{acoustic},
if  the particle goes through the bar transversally, 
in order to depose $100$~mK in the bar
we need a value
of $dE/dx$ larger by a factor $\sqrt{20}\simeq 4.5$ compared to the
longitudinal case, i.e.
\be\label{dEdx2}
\frac{dE}{dx}\simeq 90\, \frac{\rm GeV}{\rm cm}\, 
\ee
in Aluminum.
For a transverse length $2R=60$ cm this  gives a loss of
$5.4\times 10^3$~GeV in the bar, very similar to the longitudinal
case, plus about the same loss in the atmosphere, so
again the initial particle energy should have been 
at least $10^4$~GeV.

A more important modifications is  due to the fact that
the acceptance $A_T$ of the detector  is now much bigger, and
\eq{EV} becomes
\be\label{EV2}
E_{\rm tot}\simeq E\,\frac{4\pi r^2}{A_T}\simeq 
5\times 10^{-15}\, \( \frac{E}{10^4\,\rm GeV} \)
\( \frac{r}{1\,\rm kpc} \)^2 \, \msun c^2\, .
\ee
Correspondingly, \eq{E102} is replaced by
\be\label{E1022}
1<\frac{E}{10^4\,\rm GeV} < 2\times 10^{12}
\( \frac{1\,\rm kpc}{r}\)^2\, ,
\ee
where, beside using the new acceptance, we also 
used a typical galactic distance as a
reference value for $r$.
\Eq{m107} becomes
\be\label{m1072}
m < 2\times 10^{10}
\(\frac{1\, \rm kpc}{r}\)^{5/2}\, {\rm GeV}\, .
\ee
The
flux would instead be
\be\label{F200b}
F\sim \frac{200}{\rm yr\,\times 22.6 \, {\rm m^2}\, {\rm sr}}
\simeq    3\times 10^{-11}\, {\rm cm}^{-2} {\rm s}^{-1} {\rm
  sr}^{-1}\, .
\ee
A flux of this order of magnitudes still remains very difficult to
reconcile with cosmic ray experiments (even if some uncertainty can
appear, because the experimental limits depend on the detailed interaction
properties of the particle in question). 

Furthermore, despite 
the fact that the constraint on the mass given in
\eq{m1072} is now much less tight, still there is no plausible
particle physics candidate. In particular, hypothetical  neutrons 
coming from an astrophysical source would be
stopped
in the atmosphere in about 1~km;  all  neutrons detected at see level are
produced in the collision of primaries with atmospheric
nuclei, so they cannot account for a correlated signal in the two bars.
Exotic objects like nuclearites and Q-balls are still excluded by the
experimental limits on the flux, which are 3 orders of magnitude
tighter than the value required by \eq{F200b}. More in general, any object
coming from outer space,
with an energy loss of the order of the value given in \eq{dEdx2}
and a flux at see level as in \eq{F200b}
should have been seen in  cosmic ray experiments at altitude.

In conclusion, it seems that we can safely exclude that
the coincident events  are caused by the
passage of massive particles.

\section{Conclusion}\label{sectConc}

In the EXPLORER/NAUTILUS 2001 run there is a potentially 
interesting effect, i.e., an excess of 
coincidences when the two detectors
are perpendicular to the galactic plane, with a 
correlation in energy between the events in the two bars. 
Therefore, we think that it is appropriate 
to start investigating the possibility that galactic burst sources 
could explain these observations.
It is evident from our discussion  that there is no obvious
physical explanation for these data. We have examined a number of
different possibilities, and discussed the difficulties of each.
If confirmed, a crucial clue for 
understanding this effect will be provided by the sidereal time
distribution of events, which, with sufficiently large statistics,
will be able to discriminate between a population of sources
distributed in the galactic disk, a population of 
sources located at the galactic
centers or a single close source.

Another important clue might be a time clustering of the
events, with recurring bursts on the timescale of hours, and periods of
highly enhanced activity on the scale of the month. This would suggest
a kind of activity similar to what is observed in Soft Gamma Repeaters.

New data from GW detectors (the large interferometers
 LIGO, VIRGO, GEO and TAMA and 
the bars ALLEGRO, AURIGA, EXPLORER and NAUTILUS) will soon be 
available, and hopefully the situation will be clarified in the near future.

\appendix

\section{Computation of $\theta (t), \psi (t)$}
We have seen in \eq{pola} that the response function of the detector
depends on the angle $\th$ and $\psi$. In turn, these angles depend on
sidereal time because of the rotation of the Earth. The dependence of
$\th$ on $t$ has been given in \eq{theta}. In this appendix we perform the
explicit computation, both for 
$\th (t)$ and for $\psi (t)$.

\begin{figure}
\includegraphics[width=0.5\textwidth, angle=0]{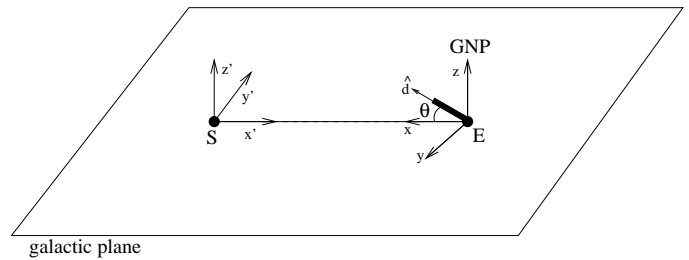}
\caption{\label{figpsi} The geometric setting discussed in appendix~A.
}
\end{figure}

For definiteness, we consider a source in the galactic plane, but the
computation is easily adapted to the most general case.
The geometric setting is shown in fig.~\ref{figpsi}. In the galactic
plane, we consider a source $S$ and the Earth (marked by $E$). On the source, we
build a reference frame with the $x'$ axis pointing toward the Earth
and the $z'$ axis in the direction
of the Galactic North Pole (GNP). Therefore the $y'$ axis lies in the
Galactic plane. The  GW propagates along the $x'$ axis and therefore
its transverse plane is $(y',z')$, and the $+$ and $\times$
polarizations are defined with respect to the $y'$ and $z'$ axes.

On the Earth we build a reference frame with the $x$ axis pointing
toward the source, the $z$ axis toward the GNP and therefore the $y$
axis in the galactic plane. We denote by $\hat{d}$ the unit vector in
the direction of the longitudinal axis of the bar. The angle between
$\hat{d}$ and the $x$ axis is $\th$. The angle $\psi$ instead
describes how the bar is rotated in the $(y,z)$ plane, i.e., with respect to
the  axes which are parallel  to the $(y',z')$
axes used to define the polarizations $+$ and $\times$. Therefore the
unit vector $\hat{d}$ can be written, in the $(x,y,z)$ frame, as
\be\label{Appdxyz}
\hat{d}=\hat{x}\cos\th +\hat{y}\sin\th\cos\psi
+\hat{z}\sin\th\sin\psi\, .
\ee
On the other hand,
the direction $\hat{d}$ of the bar is given 
by  its location on Earth 
(expressed by the latitude $l$ and longitude $L$),
its azimuthal angle $a$ and the local
sidereal time $t$. 
At the detector location, we define a new 
coordinate system with axes
$(\hat{E},\hat{N}_L,\hat{Z})$, 
where $\hat{E}$ is the east direction, $\hat{N}_L$ 
the local north (both vectors are therefore in the  plan tangent to the 
Earth surface) and $\hat{Z}$ points toward the zenith, so
$-\hat{Z}$ points toward the Earth center. 
In this frame the bar direction is given by
\begin{equation}
\hat{d}=\hat{E}\sin a +\hat{N}_L\cos a\, .
\end{equation}

\begin{figure}
\includegraphics[width=0.3\textwidth, angle=0]{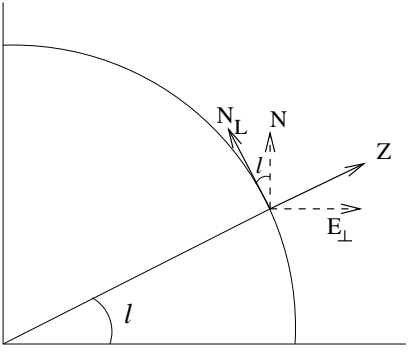}
\caption{\label{figNLN} The rotation between the coordinates systems 
$(\hat{E},\hat{N}_L,\hat{Z})$ and
$(\hat{E},\hat{N},\hat{E}_\perp)$. The unit vector $\hat{E}$
is perpendicular to the plane of the figure, pointing downward.
}
\end{figure}

The position of a source in the sky, with respect to the Earth, 
is given  in equatorial
coordinates, i.e., in the basis
$(\hat{\gamma} , \hat{\gamma}_\perp , \hat{N})$, 
where $\hat{N}$ points to the celestial north pole, 
$\hat{\gamma}$ to the vernal point 
and $\hat{\gamma}_\perp=\hat{N}\times\hat{\gamma}$. 
In this basis a unit vector $\hat{n}$ pointing toward a source
is expressed
in terms of the angles $(\alpha,\delta)$ as
\begin{equation}\label{Appn}
\hat{n}=
\hat{\gamma}\cos\alpha\cos\delta+
\hat{\gamma}_{\perp}\sin\alpha\cos\delta+
\hat{N}\sin\delta\ .
\end{equation}
In order to express $\hat{d}$ in equatorial coordinates, we first 
pass from the basis 
$(\hat{E},\hat{N}_L,\hat{Z})$
to the basis $(\hat{E},\hat{N},\hat{E}_\perp)$ 
where $\hat{E}_\perp=\hat{E}\times\hat{N}$. 
This is done with a 
rotation by an angle $l$ around the direction $\hat{E}$,
see fig.~\ref{figNLN}.
Next we perform a  rotation by an angle
$\Omega t$ around $\hat{N}$. This brings us to 
the  basis $(\hat{\gamma}_\perp ,  \hat{N},\hat{\gamma} )$.
(If instead of working with local sidereal 
time $t$ we want to use Greenwich sidereal time, we must 
rather rotate by an angle $\Omega t_{\rm Green.}+ L$). Therefore in the
basis $(\hat{\gamma}_\perp ,  \hat{N},\hat{\gamma} )$
 the components of $\hat{d}$ are given by

\be
\pmatrix{\cos\Omega t&0&\sin\Omega t\cr 0&1&0\cr
-\sin\Omega t&0&\cos\Omega  t}
\pmatrix{1&0 &0 \cr 0&\cos l&\sin l\cr
0&-\sin l&\cos l}\pmatrix{\sin a \cr \cos a \cr 0},
\ee
or, explicitly,
\bees\label{Appd}
\hat{d}&=& \hat{\gamma}\(  -\sin a\sin\O t-\cos a\sin l\cos\O t\) +\\
&+&\hat{\gamma}_{\perp}\(  \sin a\cos\O t-\cos a\sin l\sin\O t\) +
\hat{N}\cos a\sin l\nn\, .
\ees
Using \eqs{Appn}{Appd} we can now compute $\cos\th (t)$ from
$\cos\th (t) = \hat{n}\cdot\hat{d}$, and we get \eq{theta}.

\begin{figure}
\includegraphics[width=0.3\textwidth, angle=0]{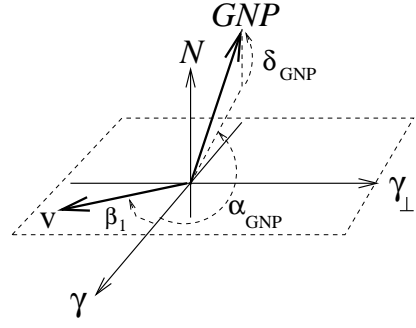}
\caption{\label{figNGNP} The vector
  $\hat{v}=\hat{N}\times\hat{z}$, where $\hat{z}$ point in the
  direction of the Galactic North Pole. The vector $\hat{v}$ is in the
$(\g ,\g_{\perp})$ plane.
}
\end{figure}

In order to compute $\psi (t)$ for a source in the galactic plane
we rather need  to transform \eq{Appd}, which gives the $t$-dependence
of $\hat{d}$, into the $(\hat{x},\hat{y},\hat{z})$ frame,
and then we can read $\psi$ by comparing with \eq{Appdxyz}.
Therefore we  must find the rotation that brings the frame
$(\hat{\gamma} , \hat{\gamma}_\perp , \hat{N})$ into the frame 
$(\hat{x},\hat{y},\hat{z})$.
This can be done in steps,  introducing first a vector
$v=\hat{N}\times\hat{z}$. The unit vector
$\hat{v}=v/\parallel v \parallel$ is given by
\begin{equation}
\hat{v}=-\hat{\gamma}\, \sin
\a_{\rm GNP}+\hat{\g}_{\perp}\cos\a_{\rm GNP}\, ,
\end{equation}
where $\a_{\rm GNP}$ is the right ascension of the Galactic North
Pole (recall that $\hat{z}$ points in the direction of the GNP).
The vector $\hat{v}$ lies in the plane
$(\hat{\g},\hat{\g}_{\perp})$, see fig.~\ref{figNGNP}, and makes an
angle $\b_1$ with the $\hat{\g}$ axis, with
\begin{equation}
 \beta_1=\arccos(\hat{\gamma}\cdot\hat{v})=\arccos(-\sin\a_{\rm GNP})\ .
\end{equation}

Then we can pass from the frame 
$(\hat{\gamma} , \hat{\gamma}_\perp , \hat{N})$ to the frame
$(\hat{v},\hat{v}_\perp,\hat{N})$, 
where $\hat{v}_\perp=\hat{N}\times\hat{v}$, performing
a rotation around $\hat{N}$ by an angle $\b_1$.
Next we  rotate the $\hat{N}$ axis into the $\hat{z}$ axis, performing
a rotation around $\hat{v}$ by an angle
\begin{equation}
\beta_2=\arccos(\hat{N}\cdot\hat{z})=\arccos(\sin\delta_{GNP})\ .
\end{equation}
This rotation leads into the basis
$(\hat{v},\hat{v'}_\perp,\hat{z})$,
where $\hat{v}$ and $\hat{v'}_\perp=\hat{z}\times\hat{v}$ 
are both in the galactic plane. 

Finally, we perform a rotation in the galactic plane in order to bring
the first axis in the direction of the desired source. This is
accomplished by a rotation around $\hat{z}$ by an angle $\b_3$. If
$\hat{n}$ is the direction of the source and $(\a ,\d )$ its right
ascension and declination, then 
\begin{eqnarray}
\cos \beta_3=\hat{n}\cdot\hat{v}=
&&-\sin\a_{\rm GNP}\cos\alpha\cos\delta+\nonumber\\
&&+\cos\a_{\rm GNP}\sin\alpha\cos\delta\, .
\end{eqnarray}
In conclusion, the components of $\hat{d}$ in the  frame
$(\hat{x},\hat{y},\hat{z})$  are given by
\begin{eqnarray}
\pmatrix{d_x\cr d_y\cr d_z}&=&
\pmatrix{c_3&-s_3&0\cr s_3&c_3&0\cr0&0&1}
\pmatrix{1&0&0\cr 0&c_2&s_2\cr0&-s_2&c_2}
\nonumber\times\\
&&\times\pmatrix{c_1&-s_1&0\cr s_1&c_1&0\cr0&0&1}
\pmatrix{d_{\g}\cr d_{\g_{\perp}}\cr d_N}\, ,
\end{eqnarray}
where $c_i=\cos\b_i, s_i=\sin\b_i$, $(i=1,2,3)$ and
$d_{\g}, d_{\g_{\perp}},d_N$ are the component in the 
 $(\hat{\gamma} , \hat{\gamma}_\perp , \hat{N})$ frame.
Comparing with \eq{Appdxyz}, we then find $\psi (t)$ from
$\tan\psi =d_z/d_y$. Figs.~\ref{figpol1}--\ref{figVelPol} are
 obtained from these expressions, shifting sidereal
time in order  to obtain the mean sidereal time between the NAUTILUS
and EXPLORER locations.

\section{Microlensing of GWs}

A magnification effect could help in explaining the results, and we
have investigated whether 
the microlensing of GWs can play a role in our problem.  We find that the
answer is clearly negative, but still we consider interesting to resume in this
appendix the reasons.
 
\begin{figure}
\includegraphics[width=0.4\textwidth]{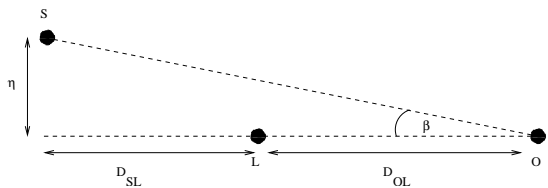}
\caption{\label{figlensing} The geometry for microlensing discussed in
the text. S,L,O denote the source, lens and observer, respectively.}
\end{figure}

Gravitational waves, just like electromagnetic waves, can be lensed by a
large mass situated between the source and the observer. 
The standard calculation of 
microlensing is performed using geometrical
optics, and in this approximation
the amplification factor ${\cal A}$ in the energy
density is (see e.g. ref.~\cite{BM})
\be\label{ampl}
{\cal A}=\frac{u^2+2}{u\sqrt{u^2+4}}\simeq\frac{1}{u}
\ee
where $u=\b /\th_E$, $\b$ is the angle of the source with respect to
the observer-lens axis (see fig.~\ref{figlensing}) and $\th_E$ is the
Einstein angle,
\be\label{thE}
\th_E =\( 2R_S\,\frac{D_{SL}}{D_{OL}(D_{SL}+D_{OL})}\)^{1/2}\, .
\ee
Here $R_S$ is the Schwarzschild radius of the lens and $D_{SL}$
and $D_{OL}$ are defined in fig.~\ref{figlensing}.
The second equality in \eq{ampl} holds when
$u\ll 1$, i.e. when the source, lens and observer are well aligned.

There is
however a crucial difference between the amplification of
electromagnetic waves and of GWs. The geometric optics approximation
holds when the reduced wavelength of the wave, $\l/(2\pi )$, is much
smaller than the typical curvature radius of the spacetime, which
in this case is given by the Schwarzschild radius of the lens, $R_S$. When
$\l/(2\pi )$ becomes of order $R_S$ diffraction effects become
important and the magnification disappears~\cite{Oha,BlM,BHa,Th2}.
When the lens is a stellar mass object, its  Schwarzschild radius is
of the order of a few kms. For visible light, therefore, the condition
$\l/(2\pi )\ll R_S$ is very well satisfied. On the contrary, the GWs
searched at resonant bars have a frequency $f\simeq 1$~kHz, and
therefore $\l/(2\pi )\sim 50\, {\rm km}\gg R_S$. This means that stellar
mass objects do not amplify GWs of a detectable frequency. To obtain
some amplification, we need a lens with a  Schwarzschild radius of at
least hundreds of kms so, within the Galaxy, the only possibility is
given by supermassive BHs. In this case $\l/(2\pi )\ll R_S$, and the
maximum amplification is given by~\cite{BlM}
\be
{\cal A}_{\rm max}\simeq 4\pi^2\frac{R_S}{\l}\, .
\ee
In other words \eq{ampl} saturates, because of diffraction effects, 
at a value of the angle $\b =\b_*$ such that
\be
\frac{\th_E}{\b_*}\simeq 4\pi^2\frac{R_S}{\l}\, .
\ee
The supermassive BH at the center of the Galaxy has $M\simeq 3\times
10^6\msun$,  and therefore it could provide a maximum 
amplification factor
\be
{\cal A}_{\rm max}\simeq 1\times 10^6\, .
\ee
Unfortunately, this factor is of no help  for
our purposes. In fact, using \eq{thE} for $\th_E$, setting
$D_{OL}\simeq 8$~kpc, which is the approximate distance to the
galactic center, and using for simplicity the approximation $D_{SL}\ll
D_{OL}$ (the result below can change at most by a factor of two without this
approximation, when the source is still in the Galaxy), we find 
that such an amplification is reached only if the source, lens and
observer are aligned within an angle $\b\leq \b_*$ with
\be
\b_* \simeq 10^{-13}\(\frac{D_{SL}}{\rm pc}\)^{1/2}
\ee
and correspondingly the distance $\eta$ between the source and the
observer-lens axis (see fig.~\ref{figlensing}) must be smaller than
\be
\eta_*\simeq 3\times 10^{-2} \(\frac{D_{SL}}{\rm pc}\)^{1/2} \rsun
\, .
\ee
This is a ridiculously small distance, and the chances of finding a
source (not to mention one hundred  sources
every  year!), which emits a GW burst 
just when it is so precisely aligned
between us and the central BH are  zero.
Smaller magnification factors could be obtained within a larger region,
but, besides the fact such events still  remain extremely unlikely, smaller
magnification factors are also of limited help because in the case of
microlensing the source must be behind the central BH, and therefore at a
distance from us $r > 8$~kpc. Clearly, to alleviate the energy
requirement in \eq{1Erad} by, say, a factor of 100 it is much easier to
assume that the source is at a smaller value of $r$, say $r\lsim
1$~kpc, rather than hiding it in an extremely 
narrow cone beyond the central BH.

Another possibility that one can consider is that there is a source of
continuous GWs, rather than of bursts, that orbits the central BH, in
such a way that at a certain moment the source is well aligned for
microlensing, and its emission is suddenly amplified. 
The idea
however does not work since, even assuming an amplification
factor $10^6$, in order to produce a burst of an apparent
energy $10^{-2}\msun c^2$, released in
a period of less than $0.1$~s (which is an upper bound on the
duration of the bursts seen by the bars) the source should  have a
continuous emission at a rate $10^{-8}\msun c^2$ in a time $\leq 0.1$~s,
corresponding to a steady rate $\geq 3\msun c^2/$yr. No continuous known 
source can emit at such a huge rate, not to mention the implausibility
of the hypothesis that the location of the
orbit  is so well fine-tuned for
performing microlensing.

In conclusion, microlensing of GWs does not appear to play a role in
our problem.

\begin{acknowledgments}

We are grateful to Venya Berezinsky, Ignazio Bombaci,
Roberto Buonanno, Luciano Burderi, Valeria Ferrari, Stefano Foffa,
Alice Gasparini,  Gian Vittorio Pallottino, Guido Pizzella, Anna Rissone
and
Riccardo Sturani for many useful discussions and to Alessandro Drago
for bringing ref.~\cite{AJK} to our attention. The work of F.D.
and M.M. is partially supported by the Fond National
Suisse.

\end{acknowledgments}


\end{document}